\documentclass[
preprint,
prd,
showpacs, 
amsmath, 
amssymb, 
floatfix,
nofootinbib,
wrapfloat
]{revtex4}
\usepackage[dvips]{graphicx}
\begin{document}

%<<<<<<<<<<<<< TITLE >>>>>>>>>>>>>>>%
\title{Close-slow analysis for
head-on collision of two black holes in higher dimensions:
Bowen-York initial data}

%<<<<<<<<<<<<< AUTHOR >>>>>>>>>>>>>>>%
\author{Hirotaka Yoshino$^{(1,2)}$}
\author{Tetsuya Shiromizu$^{(3)}$}
\author{Masaru Shibata$^{(4)}$}

%<<<<<<<<<<<<< ADDRESS >>>>>>>>>>>>>>>%

\affiliation{$^{(1)}$ Department of Physics, University of Alberta, 
Edmonton, Alberta, Canada T6G 2G7}

\affiliation{$^{(2)}$ Graduate School of Science and Engineering, 
Waseda University, Tokyo 169-8555, Japan}

\affiliation{$^{(3)}$ Department of Physics, Tokyo Institute of Technology, Tokyo 152-8551, Japan}

\affiliation{$^{(4)}$ Graduate School of Arts and Sciences, University of Tokyo, Komaba, Meguro, Tokyo 
153-8902, Japan}

%<<<<<<<<<<<<< DATE >>>>>>>>>>>>>>>%
\date{October 23, 2006; revised November 28, 2006; published December 20, 2006}

%======================================%
%<<<<<<<<<<<<< ABSTRACT >>>>>>>>>>>>>>>%
%======================================%
\begin{abstract}
Scenarios of large extra dimensions have enhanced the importance for
the study of black holes in higher dimensions.  In this paper, we
analyze an axisymmetric system of two black holes. 
Specifically, the
Bowen-York method is generalized for higher dimensions 
in order to calculate the initial data 
for head-on collision of two equal-mass black holes. Then, the initial
data are evolved adopting the close-slow approximation to study
gravitational waves emitted during the collision. We derive an empirical
formula for radiation efficiency, which depends weakly on the
dimensionality. Possible implications of our results for the black 
hole formation in particle colliders are discussed. 
\end{abstract}

%<<<<<<<<<<<<< PACS NUMBER >>>>>>>>>>>>>>>%
\maketitle

%======================================%
%<<<<<<<<<<<< SECTION I  >>>>>>>>>>>>>>%
%======================================%

\section{Introduction}

Motivated by scenarios of large extra dimensions~\cite{ADD98, RS99-1,
RS99-2}, the study of black holes in higher dimensions has attracted
attention. In these scenarios, extradimensional effects play an
important role for the properties of black holes with radius smaller
than the size of extra dimensions. 
An interesting phenomena is mini--black hole
production in planned accelerators. If the Planck energy is of
$O$(TeV) as suggested in scenarios of ~\cite{ADD98, RS99-1}, phenomena associated 
with quantum gravity will be observed at the CERN Large Hadron Collider
(LHC) through black hole production~\cite{BHUA, reviews}.
At the LHC, the produced black hole is expected to settle
down to a higher-dimensional Kerr black hole after emission of
gravitational waves. The formed black hole will be subsequently
evaporated by the Hawking radiation, for which high-energy particles
are emitted and may be detected. To know the
parameters of the formed Kerr black hole which determines the feature
of the Hawking radiation observed, studies for gravitational radiation
are an important issue. 

In this paper, we study axisymmetric collision of two black holes in
higher dimensions as a first step for the investigations on the 
black holes formed in particle colliders
\footnote{Here, the incoming black holes are regarded as substitutes for the
incoming particles. We do not consider the collision of produced black holes, 
since they are evaporated within the time
scale of $10^{-27}$ s but only one black hole is produced per 1 s.}.  
This is an extension of our previous 
study~\cite{YSS05} of the higher-dimensional Brill-Lindquist initial
data and its temporal evolution by the close-limit approximation.  In
four dimensions, there is the well-known Bowen-York method~\cite{BY80}
for generating initial data of several moving black holes.  We
generalize this method for higher dimensions.  Similarly to the
four-dimensional case, the conformally transformed extrinsic curvature
is given analytically and the conformal factor should be calculated
numerically.  We perform numerical calculation for the conformal
factor in the case of the head-on collision of two equal-mass black
holes and then determine the location of apparent horizon (AH) that
encloses two black holes (the common AH). We clarify the parameter
space for common AH formation. 

Then, the initial data is evolved adopting the close-slow
approximation. In this approximation, the distance between two black
holes and the linear momentum of each black hole are regarded as small
parameters compared to the black hole size and the gravitational mass
of the system, respectively.  In this case, the system may be
considered to be a perturbed single Schwarzschild black hole.  In the
four-dimensional case, the results in the close-slow approximation
agree with those in the fully nonlinear analysis (see \cite{PP94,
BAABPPS97} and references in \cite{YSS05}), and hence, it is natural
to expect that this is also the case in the higher-dimensional case. 
We provide a formula for the radiation efficiency (i.e., the ratio
of the radiated energy to the system energy) and discuss their
dependence on the dimensionality. 

The purpose of this paper is twofold.  One is to present a formulation
for providing the initial condition of colliding black holes in higher
dimensions. Gravitational waves at the collision of two black holes are
accurately computed only by a fully nonlinear numerical relativity
simulation. Although such a simulation has not been done yet for
higher-dimensional spacetime, several groups have developed robust techniques
for numerically computing merger of binary black holes in four
dimensions \cite{numerical-relativity}. The methods used in four
dimensions can be applied for higher-dimensional problems and hence,
the simulation will be done in the near future. For such future
simulations, a method for providing initial condition we present here
will be useful.

The second purpose is to approximately evaluate gravitational
radiation in the collision of two black holes in the higher-dimensional
case. Since we adopt a linear perturbative method, a
reliable result is derived only for a small parameter space. 
However, our result for such parameters will be useful for calibrating fully nonlinear
results in the near future. 

This paper is organized as follows.  In the next section, we derive a
generalized Bowen-York formulation~\cite{BY80} for higher-dimensional
space. Applying the formulation, we construct the initial data for
head-on collision of two equal-mass black holes in Sec. III. The ADM
mass and the common AH are analyzed.  In Sec. IV, we evolve the
initial data adopting the close-slow approximation.  The master
variable of the linear perturbation around a $D$-dimensional
Schwarzschild black hole is calculated numerically and the formula for
radiation efficiency is derived. The radiation efficiency is shown to
depend weakly on the dimensionality.  Sec. V is devoted to a
discussion on possible implication of our results for the radiation
efficiency in the particle collision. In Appendix A, we present some
solutions of the extrinsic curvature in the generalized Bowen-York
formulation that were not introduced in Sec. II. In Appendix B, 
the method for imposing the initial condition for the master
variable is explained.

%======================================%
%<<<<<<<<<<<< SECTION II  >>>>>>>>>>>>>>%
%======================================%
\section{Higher-dimensional Bowen-York method}

In this section we generalize the Bowen-York formulation for higher 
dimension space and present a method for generating initial condition 
of $N$ black holes with linear momenta. 

\subsection{Formulation}

Let $\Sigma (h_{\mu\nu}, K_{\mu\nu})$ be a $(D-1)$-dimensional
spacelike hypersurface $\Sigma$ with the metric $h_{\mu\nu}$
and the extrinsic curvature $K_{\mu\nu}$ in a $D$-dimensional spacetime. 
We introduce a number $n=D-2$. The Hamiltonian and momentum constraints
are
\begin{equation}
{}^{(n+1)}R-K^{\mu\nu}K_{\mu\nu}+K^2=0,
\end{equation}
\begin{equation}
\nabla^\mu(K_{\mu\nu}-h_{\mu\nu}K)=0,
\end{equation}
where ${}^{(n+1)}R$ denotes the Ricci scalar of $\Sigma$
and $\nabla^\mu$ is the covariant derivative with respect to $h_{\mu\nu}$.
Introducing $\widehat{h}_{\mu\nu}$ defined by 
\begin{equation} 
\widehat{h}_{\mu\nu}=\varPsi^{-4/(n-1)}h_{\mu\nu},
\end{equation}
the Hamiltonian and momentum constraints are rewritten as
\begin{equation}
\widehat{\nabla}^2\varPsi
=\frac{n-1}{4n}
\left[ {}^{(n+1)}\widehat{R}\varPsi
-\varPsi^{(n+3)/(n-1)}(K^{\mu\nu}K_{\mu\nu}-K^2)\right],
\end{equation}
\begin{equation}
\widehat{\nabla}_\nu\left(\varPsi^{2(n+1)/(n-1)}K^\nu_\mu\right)
-\varPsi^{2(n+1)/(n-1)}\widehat{\nabla}_\mu K
-\frac{2}{n-1}K\varPsi^{(n+3)/(n-1)}\widehat{\nabla}_\mu\varPsi=0,
\end{equation}
where ${}^{(n+1)}\widehat{R}$ and $\widehat{\nabla}_\mu$ denote the
Ricci scalar and the covariant derivative with respect to
$\widehat{h}_{\mu\nu}$, respectively.  Here raising and lowering the index
are done with $h_{\mu\nu}$.

In the following, we assume the conformal flatness on $\Sigma$,
$\widehat{h}_{\mu\nu}=\delta_{\mu\nu}$, and impose the maximal slicing
condition, $K=0$.  A weighted extrinsic curvature is defined by
$\widehat{K}_{\mu\nu}=\varPsi^2K_{\mu\nu}$ and hereafter its index is
raised and lowered by $\delta_{\mu\nu}$ (i.e.,
$\widehat{K}^{\mu}_\nu=\varPsi^{2(n+1)/(n-1)}K^\mu_\nu$ and
$\widehat{K}^{\mu\nu}=\varPsi^{2(n+3)/(n-1)}K^{\mu\nu}$).  Then the
Hamiltonian and momentum constraints become 
\begin{equation}
\nabla_{\rm f}^2\varPsi=-\frac{n-1}{4n}\widehat{K}_{\mu\nu}\widehat{K}^{\mu\nu}\varPsi^{-(3n+1)/(n-1)},
\label{Hamiltonian}
\end{equation}
\begin{equation}
\partial^\mu\widehat{K}_{\mu\nu}=0,
\label{momentum}
\end{equation}
where $\partial_\mu$ denotes the ordinary derivative 
with respect to the Cartesian coordinate $(x^\mu)$
and $\nabla_{\rm f}^2=\partial^\mu\partial_\mu$.
Following Bowen and York~\cite{BY80}, we assume that 
$\widehat{K}_{\mu\nu}$ does not have the tensor mode 
and thus takes the following form:
\begin{equation}
\widehat{K}_{\mu\nu}=\partial_\mu
W_\nu+\partial_\nu W_\mu-\frac{2}{n+1}\delta_{\mu\nu}\partial_\rho W^\rho.
\end{equation}
Substituting this formula into the momentum constraint \eqref{momentum}, we obtain
\begin{equation}
\nabla_{\rm f}^2W_\mu+\frac{n-1}{n+1}\partial_\mu\partial_\nu W^\nu=0.
\end{equation}
Introducing auxiliary functions $B_\mu$ and $\chi$,  we decompose $W_\mu$ as 
\begin{equation}
W_\mu=\frac{3n+1}{n-1}B_\mu-\left(\partial_\mu\chi+x^\nu\partial_\mu
B_\nu\right).
\end{equation}
Then the equation becomes
\begin{equation}
0=\frac{3n+1}{n-1}\nabla_{\rm f}^2B_\mu
-\frac{2n}{n+1}\partial_\mu\nabla_{\rm f}^2\chi
-x^\nu\partial_\mu\nabla_{\rm f}^2B_\nu
-\frac{n-1}{n+1}\partial_\mu\left(x^\nu\nabla_{\rm f}^2B_\nu\right).
\end{equation}
Hence the momentum constraint is satisfied if
\begin{align}
\nabla_{\rm f}^2B_\mu&=0,
\label{eq-Ba}\\
\nabla_{\rm f}^2\chi&=0.\label{eq-chi}
\end{align}
Since the solutions of Eqs. (\ref{eq-Ba}) and (\ref{eq-chi}) are
analytically given, solutions for $\widehat{K}_{ab}$ are easily provided. 

\subsection{$N$-black-hole solutions}

To give linear momenta of black holes,  we choose the solution
\begin{equation}
B_\mu=-\frac{2\pi GP_\mu}{n\Omega_nR^{n-1}},~~\chi=0,
\end{equation}
where $P_\mu$ is a constant vector, 
$G$ the gravitational constant, $\Omega_n$ the
$n$-dimensional area of a unit sphere, and $R= |x^\mu|$. Then, we obtain 
\begin{equation}
\widehat{K}_{\mu\nu}
=\frac{4\pi (n+1)G}{n\Omega_n R^n}
\left\{
P_\mu n_\nu+P_\nu n_\mu
+P_\rho n^\rho
\left[(n-1)n_\mu n_\nu -\delta_{\mu\nu}\right]
\right\},
\end{equation}
where $n^\mu=x^\mu/R$. This solution provides the extrinsic curvature
for one boosted black hole located at $R=0$. 
Actually, $P_\mu$ agrees with the ADM momentum: 
\begin{equation}
P_\mu=\frac{1}{8\pi G}\int_{R\to\infty} 
\left(K_{\mu\nu}n^\nu-Kn_\mu\right) dS.
\label{def-momentum}
\end{equation}

Since the momentum constraint \eqref{momentum} is a linear equation, 
we can superpose $N$ solutions. 
Denoting the locations of $N$ black holes as $x_a^\mu (a=1,...,N)$, 
a solution of the extrinsic curvature is written as 
\begin{equation}
\widehat{K}_{\mu\nu}=\sum_{a=1}^N
\frac{4\pi (n+1)G}{n\Omega_n R_a^n}
\left\{
(P_a)_\mu (n_a)_\nu+(P_a)_\nu (n_a)_\mu
+(P_a)_\rho (n_a)^\rho
\left[(n-1)(n_a)_\mu (n_a)_\nu -\delta_{\mu\nu}\right]
\right\},
\end{equation}
where $R_a=|x^\mu-x_a^\mu|$, $n_a^\mu=(x^\mu-x_a^\mu)/R_a$, and 
$(P_a)_\mu$ denotes the momentum of the $a$th black hole.  

The conformal factor $\varPsi$ is obtained by solving the Hamiltonian
constraint \eqref{Hamiltonian}. Following \cite{BB97}, 
we assume that $\varPsi$ has the following form 
\begin{equation}
\varPsi=\varPsi_{\rm BL}+\psi, 
\end{equation}
where
\begin{equation}
\varPsi_{\rm BL}\equiv
1+\frac{4\pi G}{n\Omega_n}\sum_{a=1}^N\frac{M_a}{R_a^{n-1}}
\end{equation}
and $M_a$ denotes the mass parameter of $a$th black hole. 
Then, the equation for $\psi$ becomes 
\begin{equation}
\nabla_{\rm f}^2\psi=-\frac{(n-1)}{4n}\widehat{K}_{\mu\nu}\widehat{K}^{\mu\nu}
(\varPsi_{\rm BL}+\psi)^{-(3n+1)/(n-1)}. \label{eq20}
\end{equation}
The solution in this procedure represents the so-called ``puncture'' space
with $N$ Einstein-Rosen bridges and $N+1$ asymptotically flat regions
(say, one upper sheet and $N$ lower sheets).

Since the right hand side of Eq. (\ref{eq20}) behaves like
$O(R_a^{n-1})$ for $R_a\to 0$, there is a regular solution for $\psi$,
which can be solved numerically (this fact was first pointed out in
\cite{BB97} for the four-dimensional case).  The ADM mass $M_{\rm
ADM}$ is given by 
\begin{equation}
M_{\rm ADM}=\frac{-n}{4\pi (n-1)G}\int_{R\to \infty}
\partial_\mu\varPsi n^\mu dS.
\end{equation}
Using the Gauss law, we find
\begin{eqnarray}
M_{\rm ADM}=\sum_{a=1}^N M_a+\frac{1}{16\pi G}\int_{\Sigma}
\widehat{K}^{\mu\nu}\widehat{K}_{\mu\nu}\varPsi^{-(3n+1)/(n-1)} d^{n+1}x.
\label{ADM-mass}
\end{eqnarray}
The first and second terms could be interpreted as the sum of
the mass of $N$ black holes and the kinetic energy of the black holes, 
respectively.

\section{Initial data for head-on collision of two black holes}

In this section, we present initial data of the axisymmetric
two-black-hole system following the formalism described 
in the previous section.

\subsection{Calculating conformal factor}

We introduce the Cartesian coordinate $(z, x_k), (k=1,...,n)$ for
$(n+1)$-dimensional space and write the solution of
$\widehat{K}_{\mu\nu}$ as 
\begin{equation}
\widehat{K}_{\mu\nu}=\widehat{K}_{\mu\nu}^{(+)}+\widehat{K}_{\mu\nu}^{(-)},
\end{equation}
where two black holes $(\pm)$ are located at
$(z, x_k)=(\pm z_0,0)$ and have momenta $P_\mu^{(\pm)}=(\mp P, 0)$. 
We assume that the two black holes have the same mass $M_+=M_-=M_0/2$.

The gravitational radius of a black hole of mass $M_0$ is defined by
\begin{equation}
r_h(M_0)=\left(\frac{16\pi GM_0}{n\Omega_n}\right)^{1/(n-1)}.
\end{equation}
In the case $z_0=0$, the common apparent horizon (AH) that encloses
two black holes is located at $R=R_h(M_0)$ where
\begin{equation}
R_h(M_0)=4^{-1/(n-1)}r_h(M_0). \label{Capital-RHM0}
\end{equation}
Using $R_h(M_0)$ the conformal factor in the puncture framework \cite{BB97}
is given by 
\begin{equation}
\varPsi=\varPsi_{\rm BL}+\psi,
\end{equation}
where
\begin{equation}
\varPsi_{\rm BL}=1+\frac12\left[R_h(M_0)\right]^{n-1}\left(\frac{1}{R_+^{n-1}}
+\frac{1}{R_-^{n-1}}\right).
\end{equation}

For a numerical solution of $\psi$, we introduce the cylindrical coordinate
$(z, \rho)$ where $\rho=\sqrt{\sum_{k=1}^n x_k^2}$. 
In this coordinate, the equation for $\psi$ becomes
\begin{equation}
\psi_{,\rho\rho}+\psi_{,zz}+\frac{(n-1)}{\rho}\psi_{,\rho}
+\frac{n+1}{4n}\widehat{K}_{\mu\nu}\widehat{K}^{\mu\nu}\varPsi^{-(3n+1)/(n-1)}=0.
\end{equation}
Since the system is axisymmetric and equatorial-plane symmetric, it is
sufficient to solve the equation for the domain $0\le\rho\le\rho_{\rm
max}$ and $0\le z\le z_{\rm max}$ with the boundary conditions
$\psi_{,\rho}=0$ at the $z$-axis and $\psi_{,z}=0$ at the equatorial
plane. For $R \rightarrow \infty$, $\psi$ asymptotically behaves as 
\begin{equation}
\psi\simeq\frac{4\pi G(M_{\rm ADM}-M_0)}{n\Omega_nR^{n-1}}
+O(1/R^n).
\end{equation}
Thus, we impose the so-called Robin condition, $\psi_{,R}=-(n-1)\psi/R$, 
at the outer boundary.
In the cylindrical coordinate, it is rewritten as 
\begin{equation}
\psi_{,\rho}\rho+\psi_{,z}z=-(n-1)\psi.
\end{equation}
We put the outer boundary at $\rho_{\rm max}=z_{\rm max}=5R_h(M_0)$
and solve the equation for $\psi$ using a finite difference method
with the grid number $(101\times 101)$.  Numerical computation was
carried out for the parameter space $0\le z_0/R_h(M_0)\le 1$ and $0\le
P/M_0\le 1$ with the $0.1$ interval or the 0.01 interval.

There are two sources for numerical error of the conformal factor: One is
associated with the finiteness of the grid spacing and the other with 
the finiteness of the outer boundary location.
To evaluate the numerical error
by the grid spacing, we 
took reference data with sufficiently large grid number ($401\times 401$ grids) 
and evaluated following three characteristic numerical error values:
\begin{align}
\epsilon_1&=\frac{\sum_N\left|\psi_N-\psi_N^{\rm (ref)}\right|}
{\sum_N\left|\psi_N^{\rm (ref)}\right|},
\nonumber\\
\epsilon_2&=\max\left(
{\left|\varPsi_N-\varPsi_N^{\rm (ref)}\right|}
\left/{\left|\varPsi_N^{\rm (ref)}\right|}\right.
\right),
\label{error-estimate}
\\
\epsilon_3&=\max\left({\left|\psi_N-\psi_N^{\rm (ref)}\right|}
\left/{\left|\psi_N^{\rm (ref)}\right|}\right.\right),
\nonumber
\end{align} 
where $N$ stands for the label of the grids.
Changing the grid
spacing, we confirm that $\epsilon_1$ decreases with improving
the grid resolution at second order. 
All the three error estimates
were found to be small.
With our standard choice, 
$\epsilon_2$ is $0.005$--$0.2$\% for $D=4$--$11$. Computation was also
performed changing the location of the outer boundaries from
$\rho_{\rm max}=z_{\rm max}=5R_h(M_0)$ to $20R_h(M_0)$ and the
numerical solution converges with increasing the radius of the outer
boundaries. 
In this case, $\epsilon_3$ is found to be relatively large,
about 9--17\% for $D=4$--$11$. Such large differences occur
at outer boundaries and come from the fact that the Robin condition
is an approximate boundary condition. However, because $\psi$ is
small at the outer boundary, the error of the conformal factor $\varPsi$
is small. In fact, we found that $\epsilon_2$ 
is about $0.3\%$ for $D=4$ and becomes smaller as $D$ 
is increased (less than $0.01\%$ in the case $D=11$). 
The error $\epsilon_1$ is smaller than $\epsilon_3$ by a factor for $D=4$
and is comparable to $\epsilon_2$ for $D=11$.
Because $\varPsi$ is used in the calculations of the ADM mass or the AH,
the error for these calculations is expected to have the order of $\epsilon_2$.
Hence, we consider that sufficient accuracy is obtained in our calculation.

%%%%%%%%%%%%%%%%%%%%%%%%%%%%%%%%%%%%%%%%%%%%%%%%%%%%%%%%
\begin{figure}[h]
\centering
{
\includegraphics[width=0.33\textwidth]{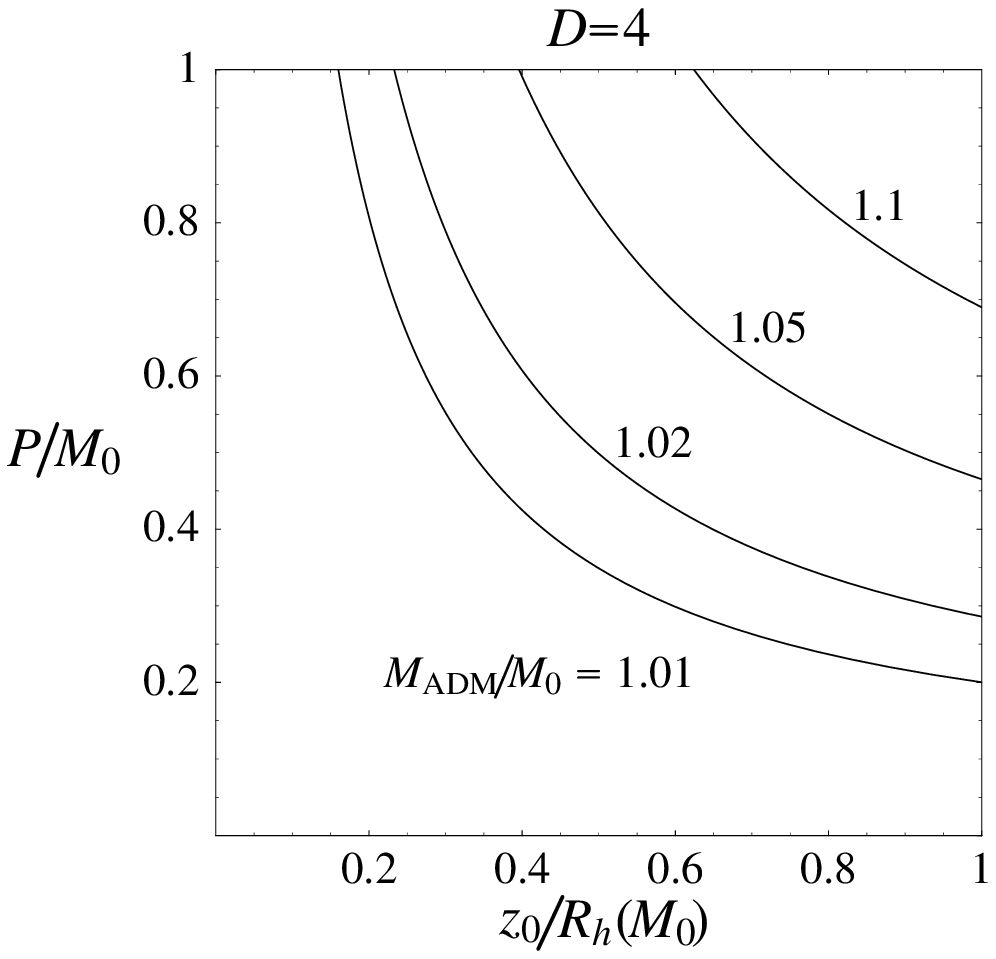}\hspace{10mm}
\includegraphics[width=0.33\textwidth]{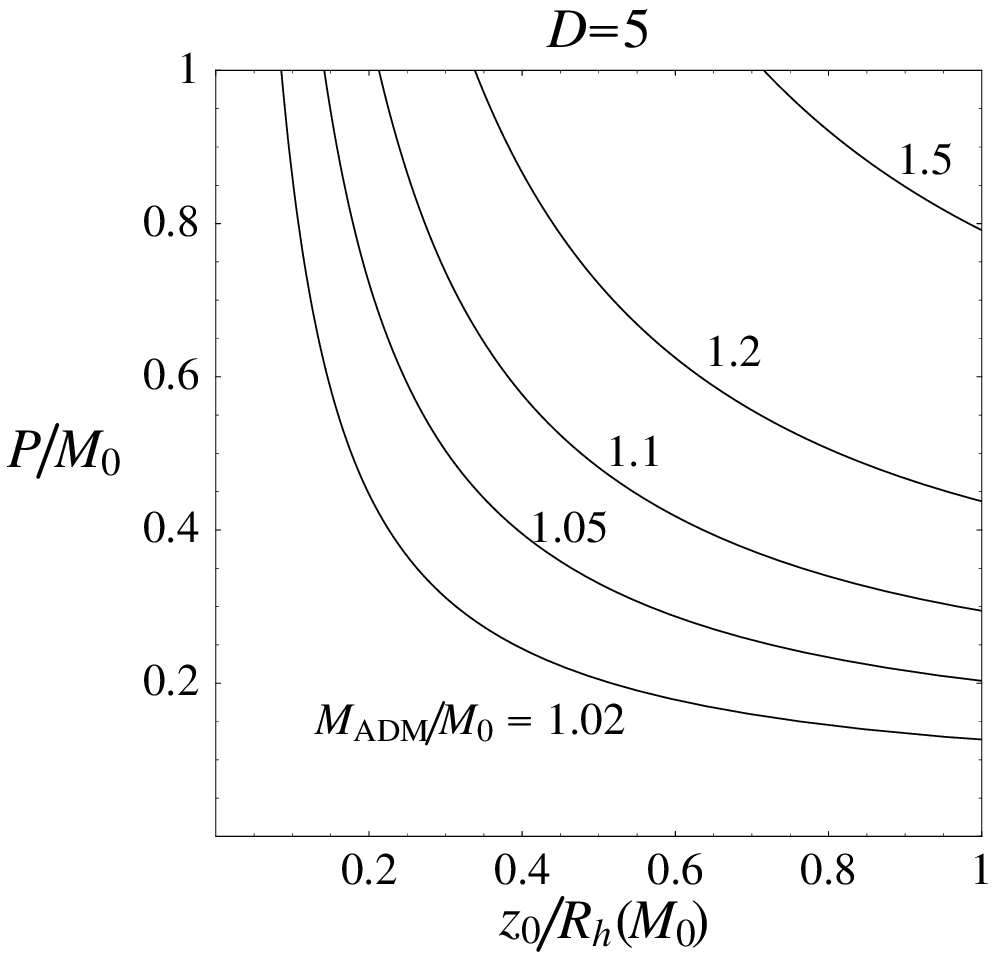}\\
\includegraphics[width=0.33\textwidth]{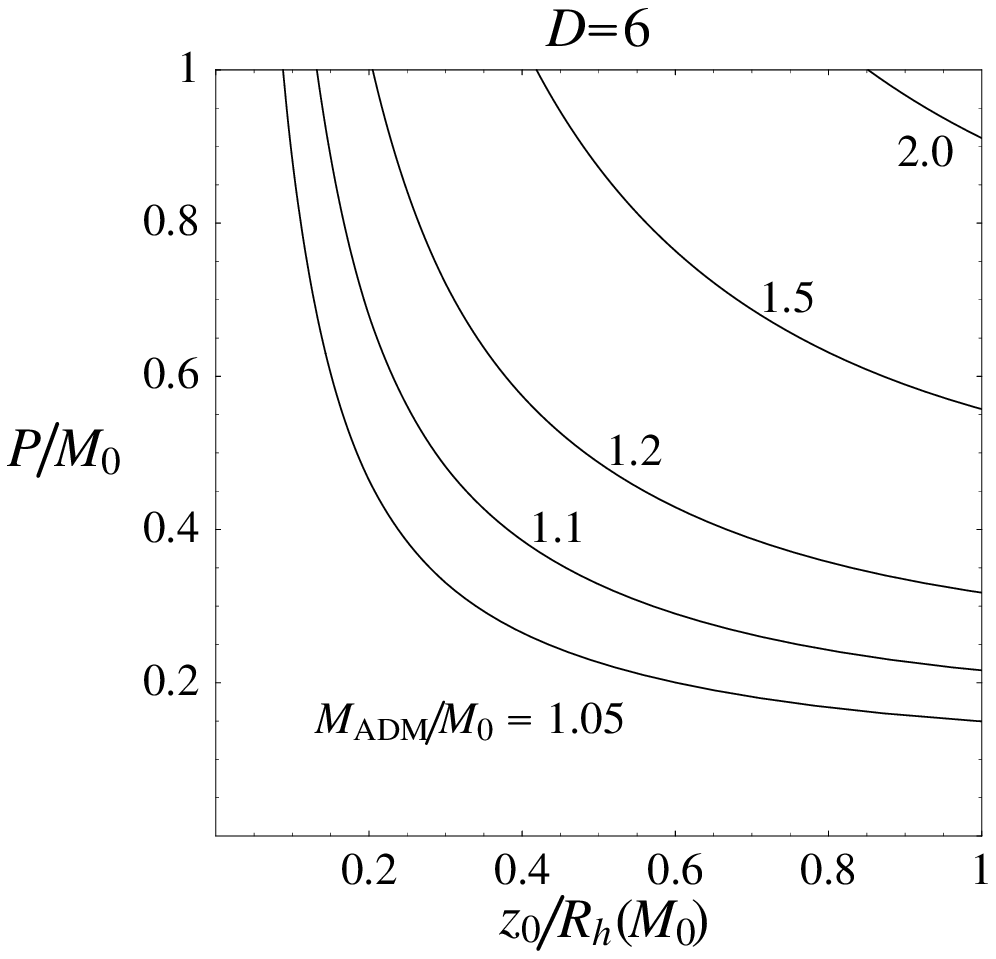}\hspace{10mm}
\includegraphics[width=0.33\textwidth]{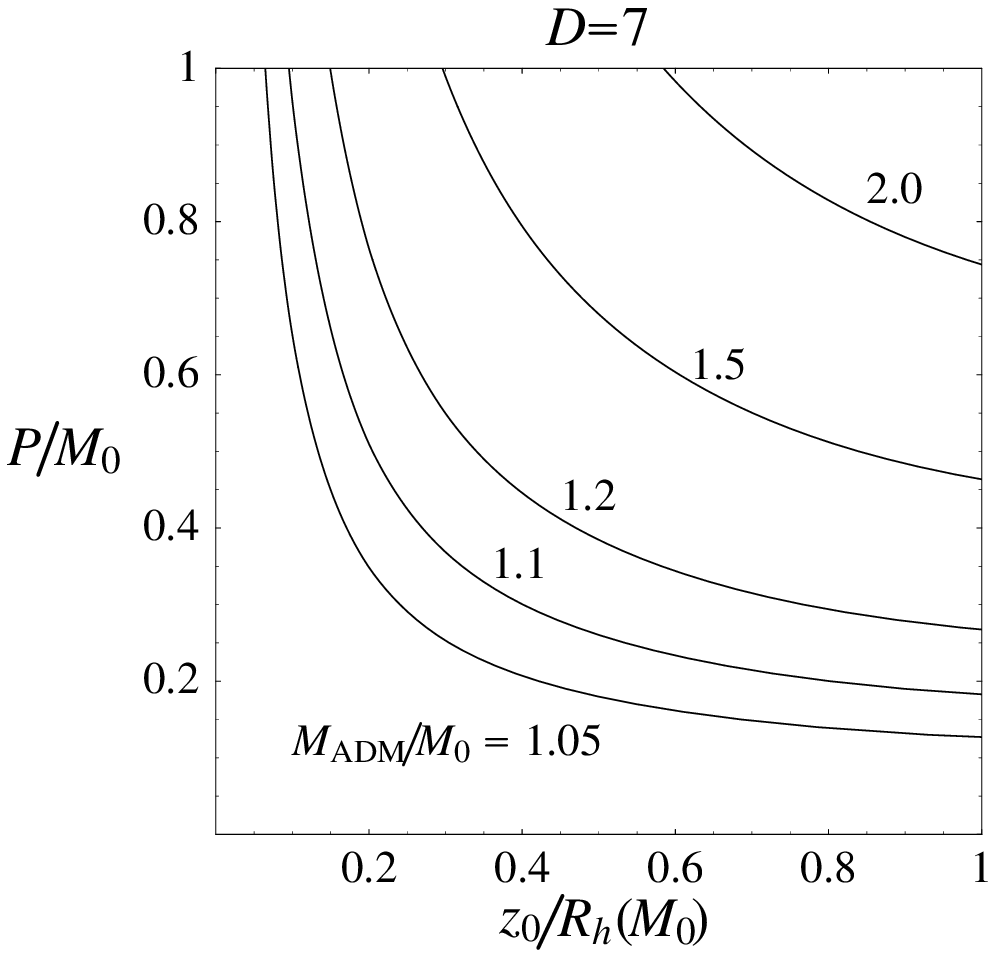}\\
\includegraphics[width=0.33\textwidth]{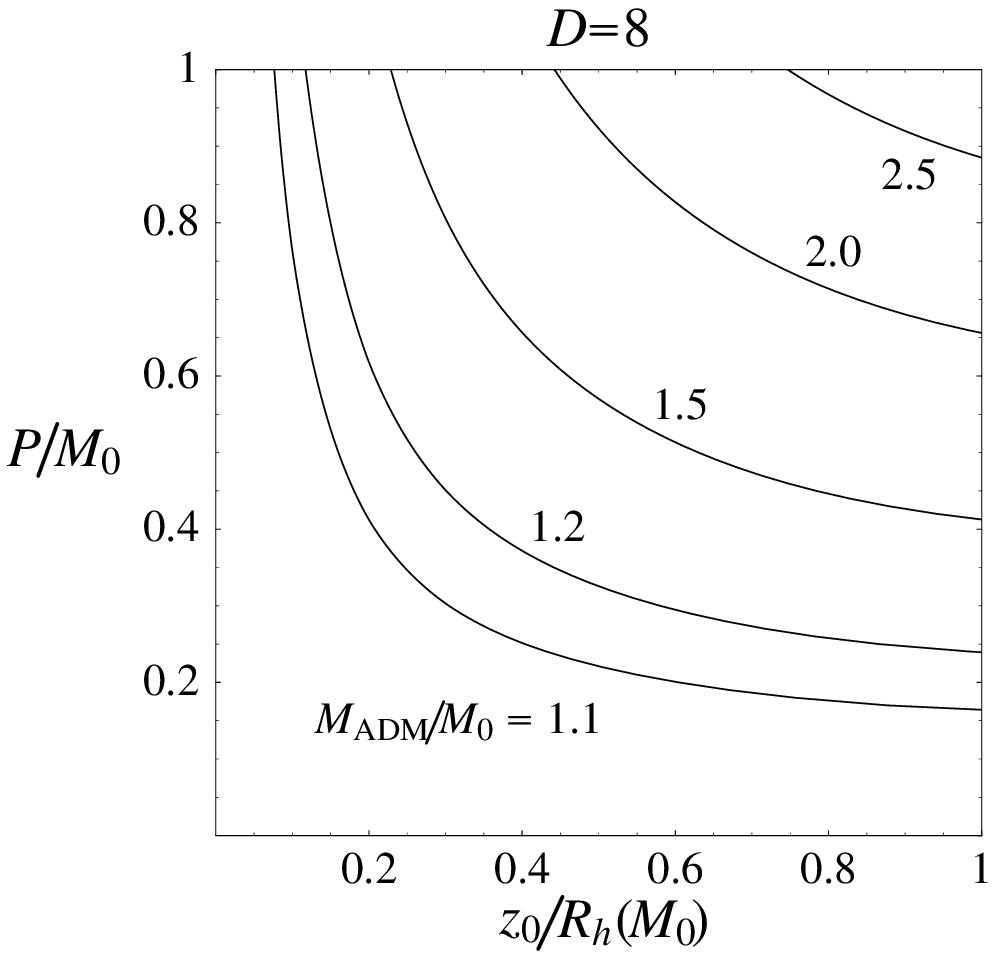}\hspace{10mm}
\includegraphics[width=0.33\textwidth]{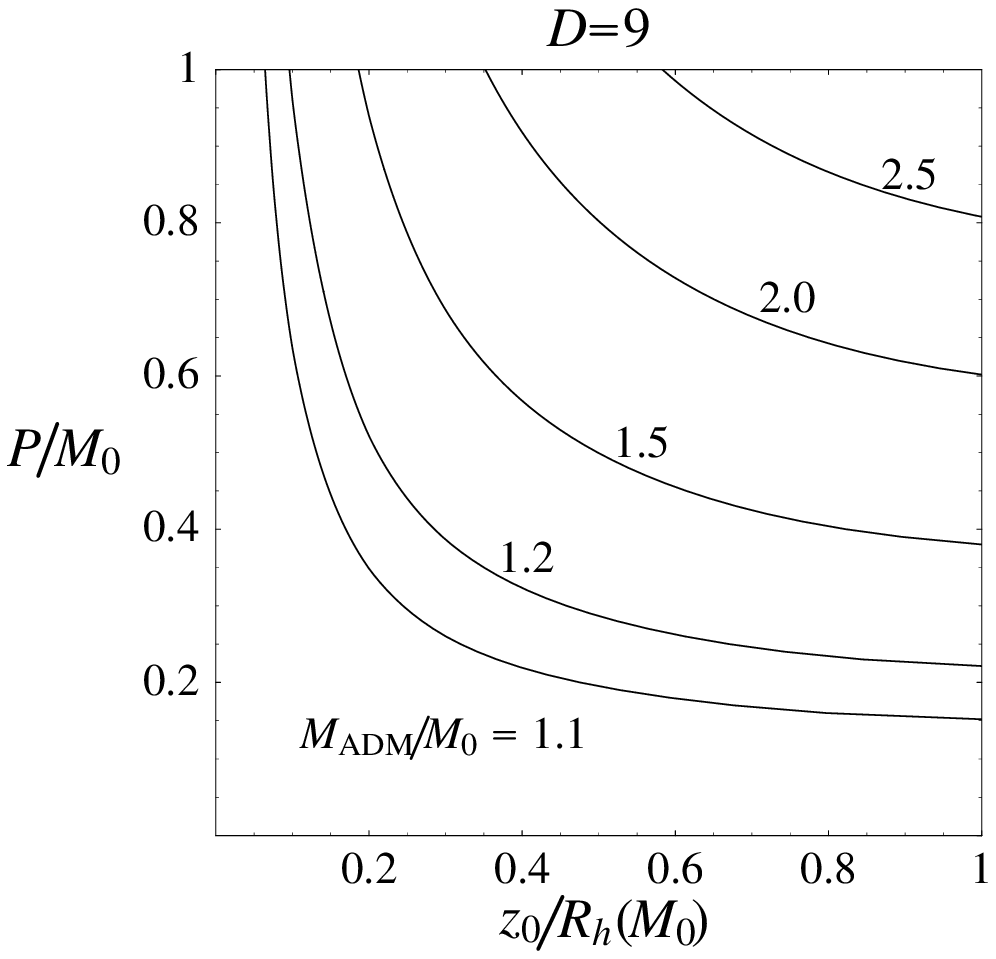}\\
\includegraphics[width=0.33\textwidth]{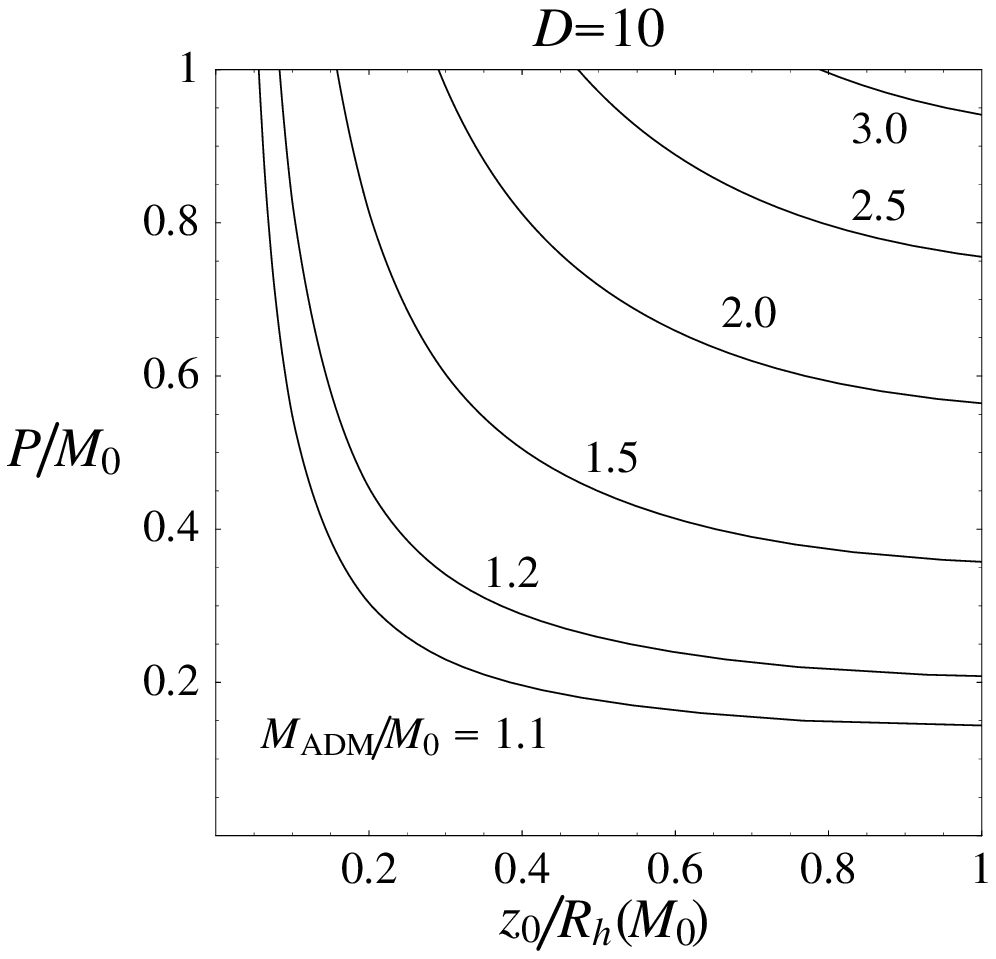}\hspace{10mm}
\includegraphics[width=0.33\textwidth]{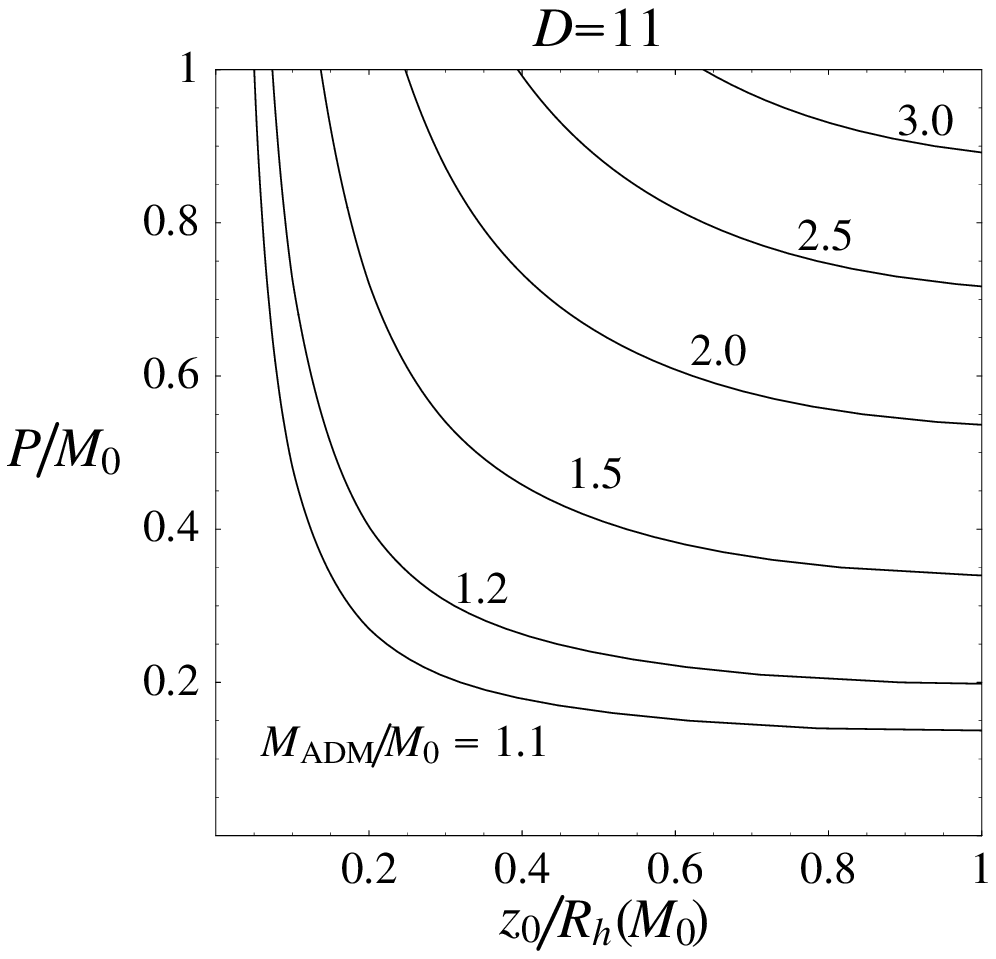}
}
\caption{The contour line for $M_{\rm ADM}/M_0$ on the $(z_0/R_h(M_0), P/M_0)$-plane.}
\label{contour-ADM}
\end{figure}
%%%%%%%%%%%%%%%%%%%%%%%%%%%%%%%%%%%%%%%%%%%%%%%%%%%%%%%%

Figure~\ref{contour-ADM} shows the contours of $M_{\rm ADM}/M_0$ on
the $(z_0/R_h(M_0), P/M_0)$-plane for $D=4$--$11$. The difference
between $M_{\rm ADM}$ and $M_0$ indicates the strength of nonlinearity
due to the right hand side of the Hamiltonian
constraint~\eqref{Hamiltonian}.  We find that the nonlinearity of the
system becomes large as $D$ is increased for fixed values of
$z_0/R_h(M_0)$ and $P/M_0$.

\subsection{Common apparent horizon}

%%%%%%%%%%%%%%%%%%%%%%%%%%%%%%%%%%%%%%%%%%%%%%%%%%%%%%%%
\begin{figure}[h]
\centering
{
\includegraphics[width=0.33\textwidth]{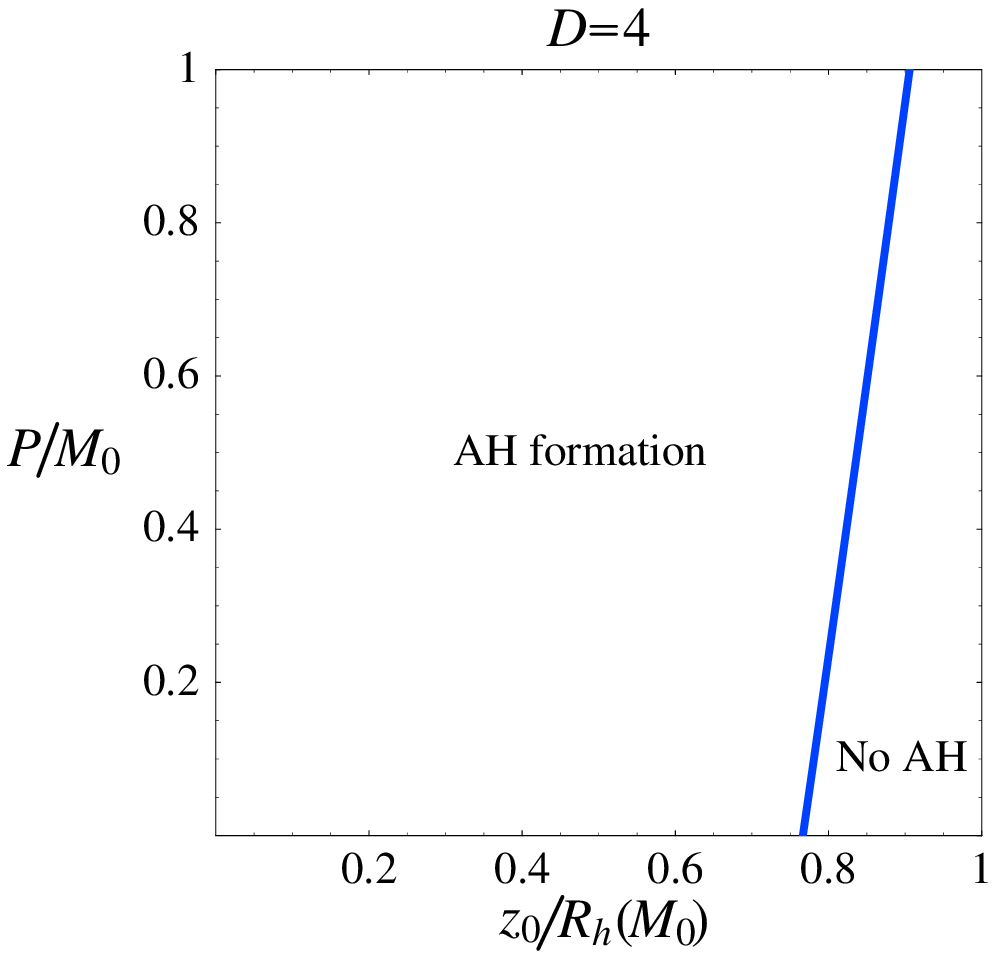}\hspace{10mm}
\includegraphics[width=0.33\textwidth]{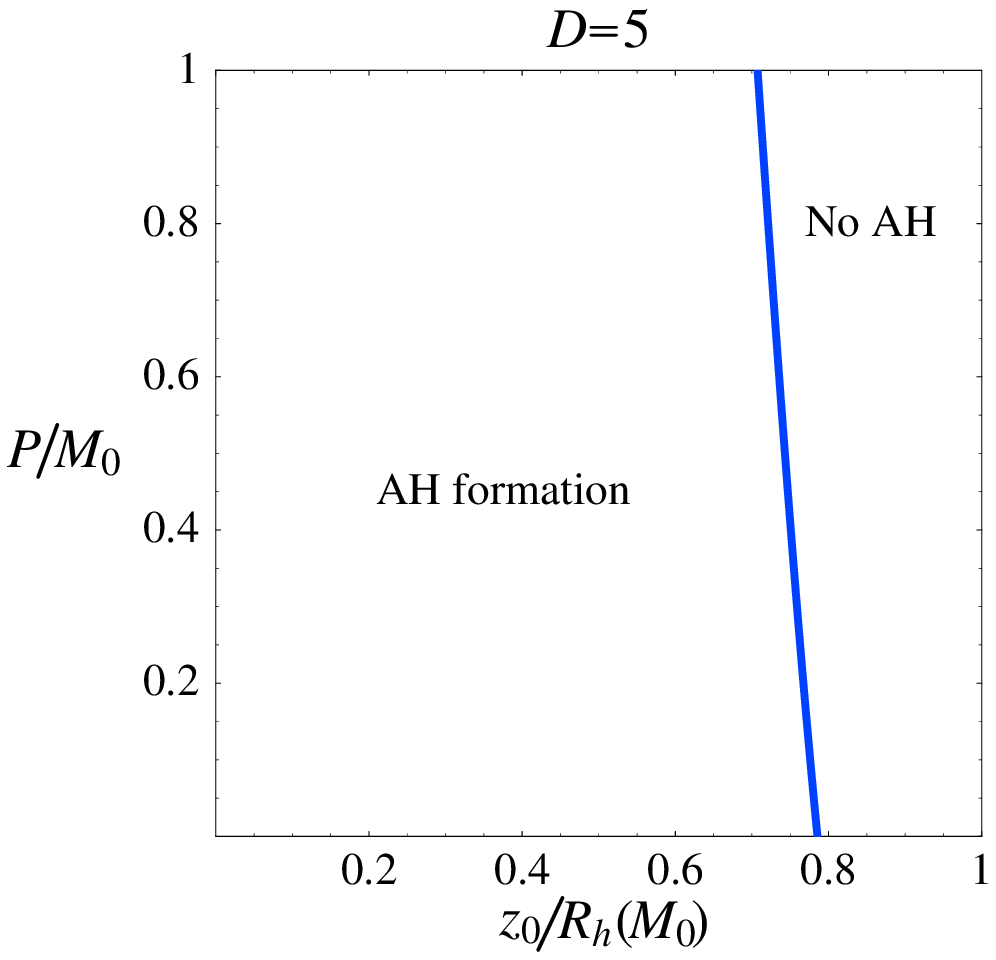}\\
\includegraphics[width=0.33\textwidth]{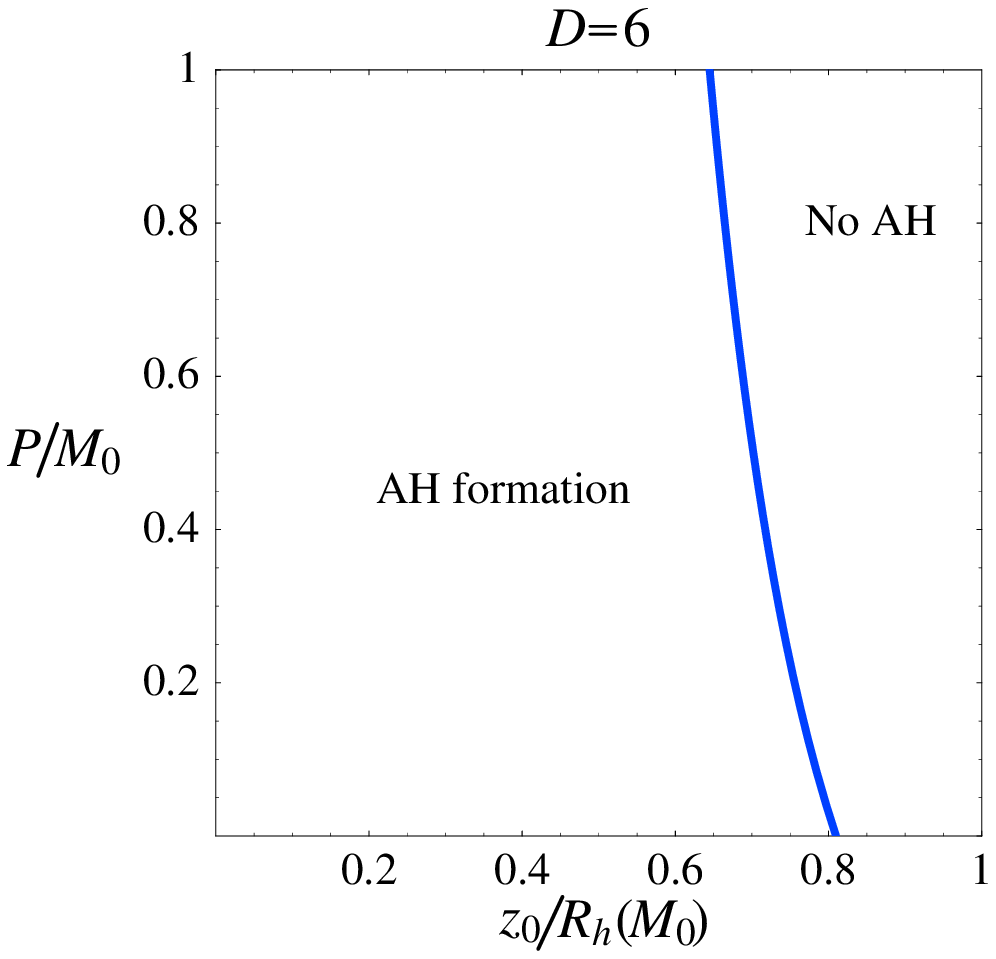}\hspace{10mm}
\includegraphics[width=0.33\textwidth]{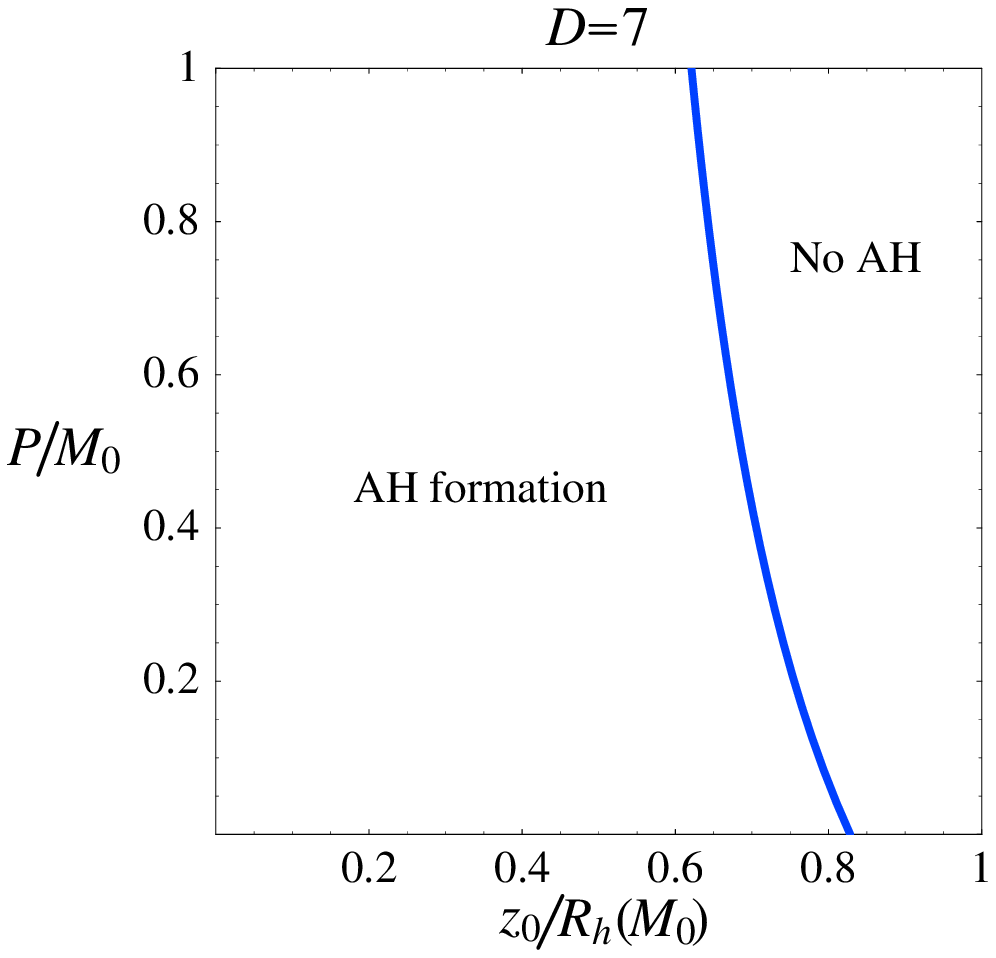}\\
\includegraphics[width=0.33\textwidth]{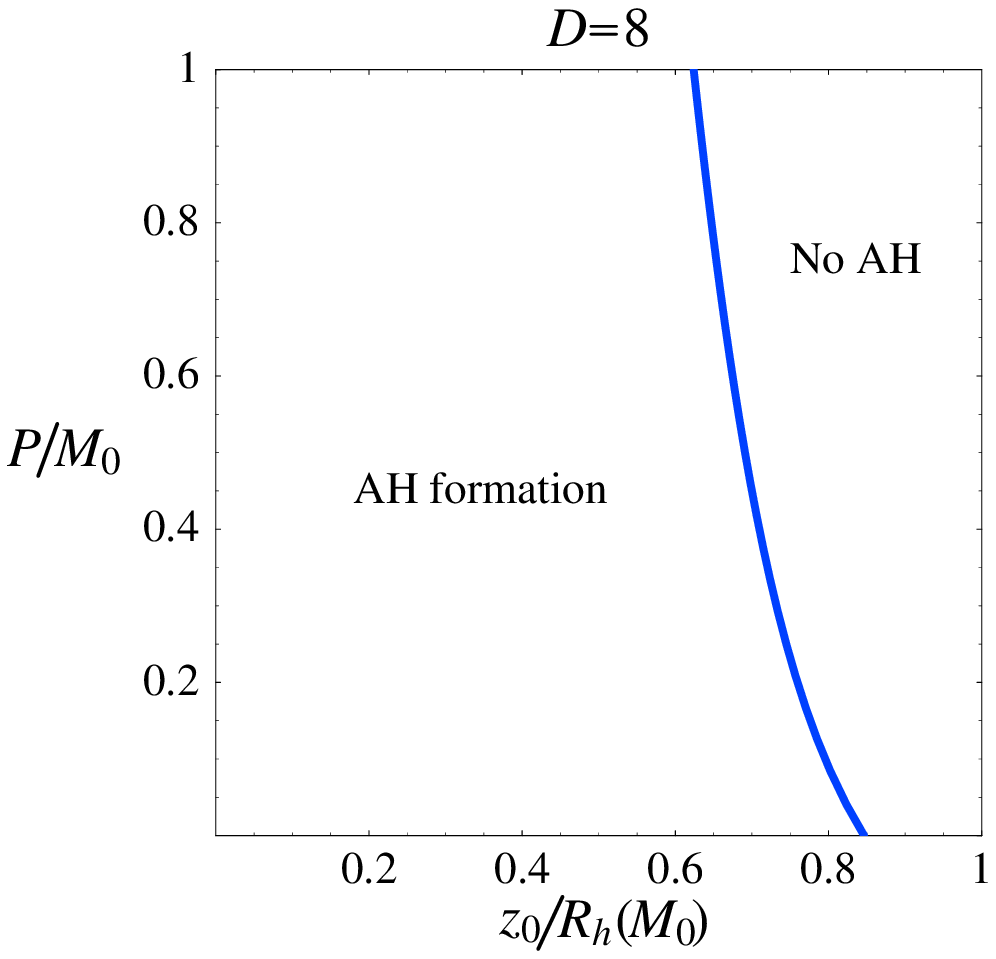}\hspace{10mm}
\includegraphics[width=0.33\textwidth]{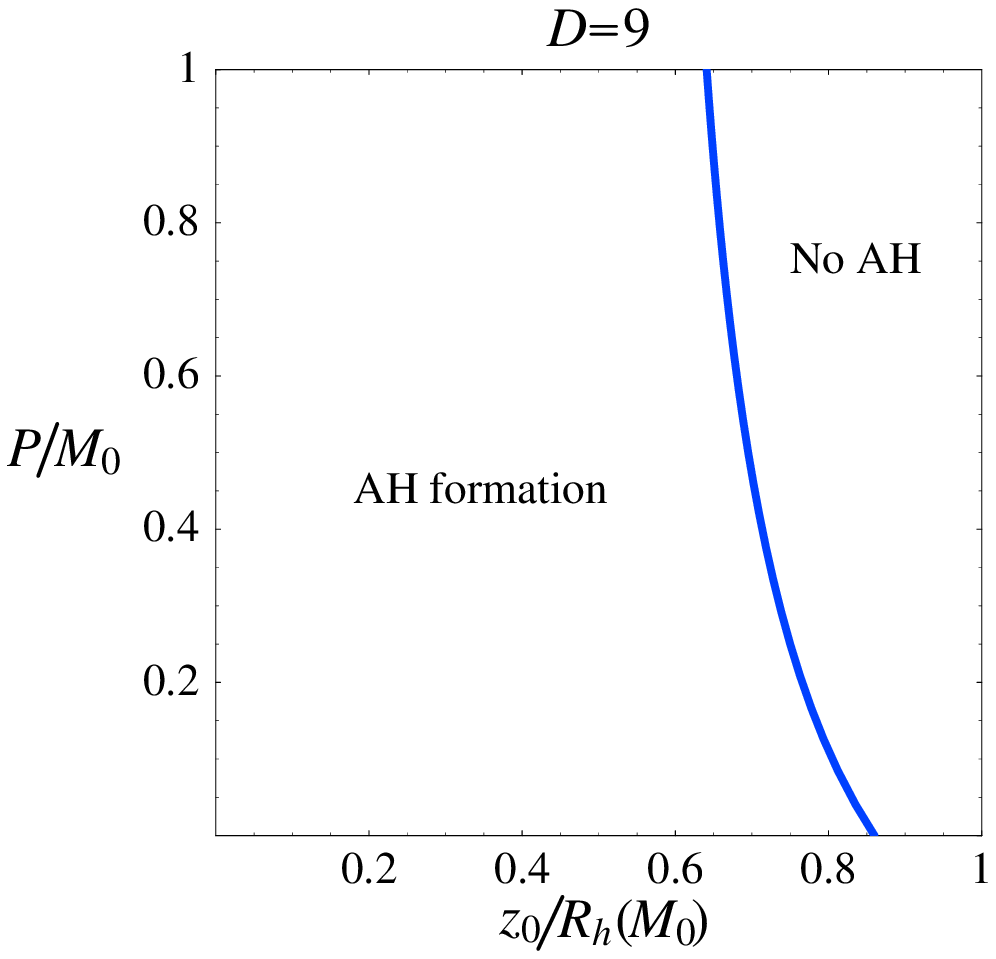}\\
\includegraphics[width=0.33\textwidth]{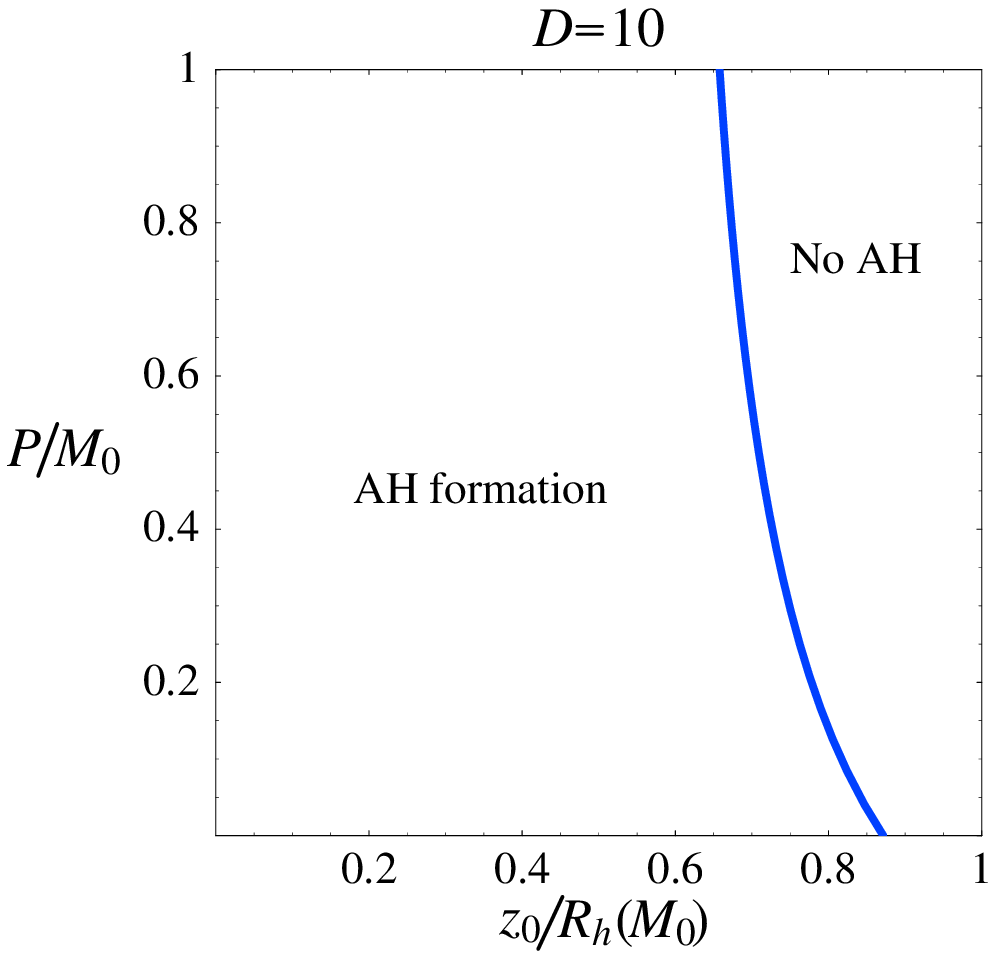}\hspace{10mm}
\includegraphics[width=0.33\textwidth]{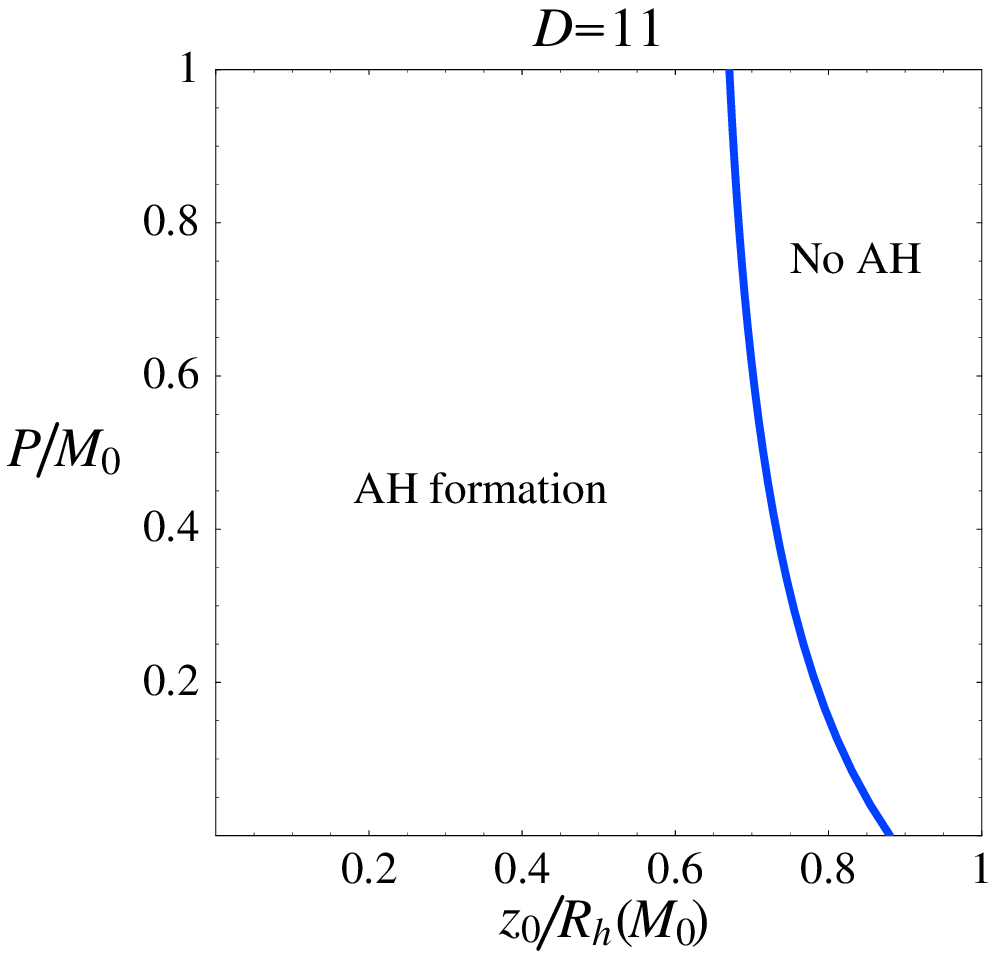}
}
\caption{The critical line for the common AH formation
on the $(z_0/R_h(M_0), P/M_0)$-plane.  }
\label{critical-AH}
\end{figure}
%%%%%%%%%%%%%%%%%%%%%%%%%%%%%%%%%%%%%%%%%%%%%%%%%%%

A common AH that covers two black holes is searched for introducing the
spherical coordinate $(R, \theta)$ by $R=\sqrt{\rho^2+z^2}$ and
$\tan\theta=z/\rho$ and assuming that the location is denoted by
$R=h(\theta)$.  Then the equation for the AH is given by
\begin{equation}
\nabla_\mu s^\mu-K+K_{\mu\nu}s^\mu s^\nu=0,
\end{equation} 
where $s^\mu$ is the unit normal to the surface:
\begin{equation}
s^\mu=\frac{\varPsi^{-2/(n-1)}}{\sqrt{1+h_{,\theta}^2/R^2}}
\left(1,-h_{,\theta}/R^2\right).
\end{equation}
Then, the equation for the AH reduces to an ordinary differential equation for 
$h(\theta)$:
\begin{multline}
h_{,\theta\theta}-n\left(\frac{2}{n-1}\frac{\varPsi_{,R}}{\varPsi}
+\frac{1}{h}\right)\left(h^2+h_{,\theta}^2\right) -
\frac{h_{,\theta}^2}{h}
+\left(\frac{2n}{n-1}\frac{\varPsi_{,\theta}}{\varPsi}
+(n-1)\cot\theta\right)h_{,\theta}
\left(1+\frac{h_{,\theta}^2}{h^2}\right)\\ -
\varPsi^{-2n/(n-1)}h\sqrt{h^2+{h_{,\theta}^2}}
\left(\widehat{K}_{RR}-2\widehat{K}_{R\theta}\frac{h_{,\theta}}{h^2}
+\widehat{K}_{\theta\theta}\frac{h_{,\theta}^2}{h^4}\right) =0.
\end{multline}
This equation is solved under the boundary conditions $h_{,\theta}=0$
at $\theta=0$ and $\pi/2$.

%%%%%%%%%%%%%%%%%%%%%%%%%%%%%%%%%%%%%%%%%%%%%%%%%%%%%%%%
\begin{figure}[tb]
\centering
{
\includegraphics[width=0.45\textwidth]{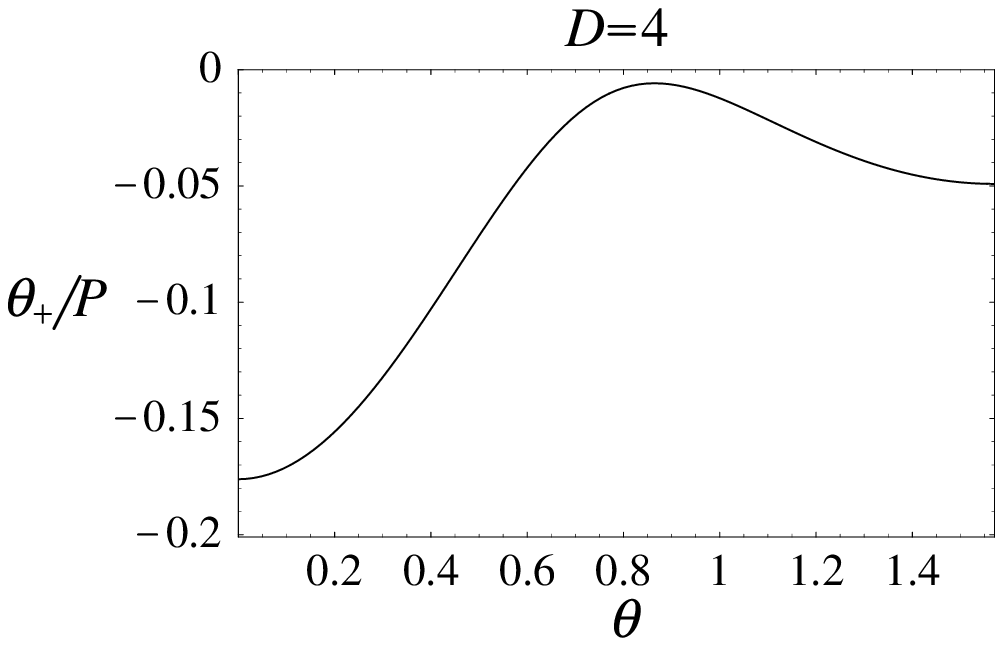}
\includegraphics[width=0.45\textwidth]{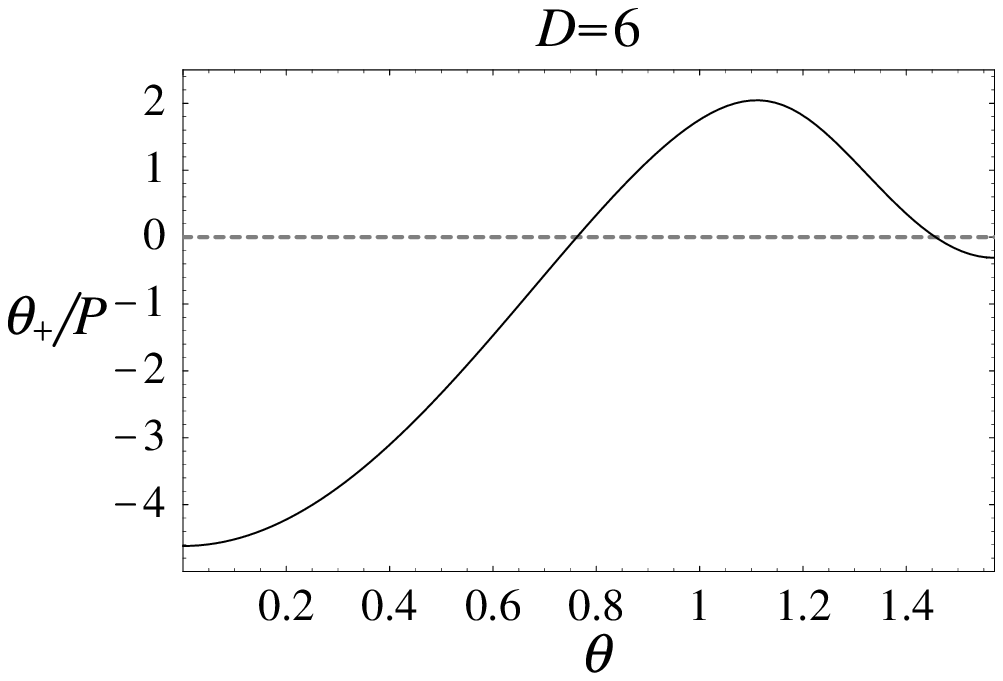}
}
\caption{The expansion $\theta_+$ of a minimal surface in the $P\ll M_0$ case
at the critical separation $z_0=z_0^{\rm (crit)}$.
The cases $D=4$ (left) and $6$ (right) are shown.
In the case $D=4$, $\theta_+$ is negative for all $\theta$ and the minimal surface
is a trapped surface. But in the case $D=6$, there is a region of $\theta$
where $\theta_+$ becomes positive, which means that the minimal surface is not
a trapped surface.}
\label{expansion}
\end{figure}
%%%%%%%%%%%%%%%%%%%%%%%%%%%%%%%%%%%%%%%%%%%%%%%%%%%%%%%%

Figure~\ref{critical-AH} shows the critical line for the common AH
formation. Irrespective of the dimensionality, the common AH is formed
if $z_0/R_h(M_0)$ is sufficiently small. It is interesting to point
out that, for $D=4$, the AH formation is more subject for larger
values of $P$ while for $D \geq 5$, the presence of $P$ tends to
prevent formation of the common AH.
This might seem strange because the kinetic energy is naturally
expected to help the AH formation. 
In order to obtain a corroboration for this result, 
we direct attention to the AH (denoted by $R=\widehat{h}(\theta)$)
at the critical separation $z_0=z_0^{\rm (crit)}$
in the case $P=0$. 
Then we add small $P\ll M_0$ without changing
the value of $z_0$. Because the change in the
conformal factor is $O(P^2/M_0^2)$ and can be ignored,
the surface $R=\widehat{h}(\theta)$ is interpreted as a minimal surface
on which $\nabla_\mu s^\mu=0$ holds. Now let us consider
the expansion on this minimal surface which
is given by $\theta_+=K_{\mu\nu}s^\mu s^\nu$. 
Figure~\ref{expansion} shows the behavior of $\theta_+/P$ 
in the four- and six-dimensional cases.
In the case $D=4$, 
we find that $\theta_+$ is negative and thus the minimal surface
is a trapped surface. This guarantees the existence
of an AH outside of the minimal surface.
Hence the motion helps AH formation in the four-dimensional case.
On the other hand, in the case $D=6$,
$\theta_+$ becomes positive on some part of the minimal surface. 
The similar behavior was found
for all $5\le D\le 11$. Therefore, the minimal surface is not a trapped surface
and the motion does not necessarily help the AH formation 
in the higher-dimensional cases.
We found numerically that the term $K_{\theta\theta}s^\theta s^\theta$
mainly contributes to the positivity of $\theta_+$. This is because for high $D$,
the AH at the critical separation is
hourglass shaped in the neighborhood of the equatorial plane 
and the value of $\widehat{h}_{,\theta}$
is quite large around $\theta\simeq 1$. It enhances the value of 
$K_{\theta\theta}s^\theta s^\theta$ which is proportional to 
${\widehat{h}_{,\theta}}^2/{\widehat{h}}^4$.

%%%%%%%%%%%%%%%%%%%%%%%%%%%%%%%%%
\begin{table}[tb]
\centering
\caption{The values of $1-M_{\rm AH}/M_{\rm ADM}$ evaluated on the AH critical line
for $P/M_0=0.0,0.5,1.0$. The unit is \%. }
\begin{ruledtabular}
\begin{tabular}{c|cccccccc}
$D$ & $4$ & $5$ & $6$ & $7$ & $8$ & $9$ & $10$ & $11$  \\
  \hline 
$P/M_0=0.0$ & $1.2$ & $3.6$ & $5.8$ & $7.2$ & $7.7$ & $7.7$ & $7.4$ & $6.9$ \\
$P/M_0=0.5$ & $2.8$ & $9.0$ & $12$ & $14$ & $16$ & $17$ & $18$ & $19$ \\
$P/M_0=1.0$ & $5.5$ & $15$ & $19$ & $21$ & $22$ & $24$ & $25$ & $26$ \\
  \end{tabular}
  \end{ruledtabular}
  \label{lower-bound}
\end{table}
%%%%%%%%%%%%%%%%%%%%%%%%%%%%%%%%%

The AH mass $M_{\rm AH}$ is determined by the AH area $A_{\rm AH}$ as 
\begin{equation}
M_{\rm AH}=\frac{n\Omega_n}{16\pi G}
\left(\frac{A_{\rm AH}}{\Omega_n}\right)^{1/n}. 
\end{equation} 
The area theorem of black hole constrains that the AH mass never 
decrease \footnote{Since the proof of the area theorem is not sensitive 
to the spacetime dimension, it holds also in the current 
system as long as naked singularities do not exist.}. 
Thus, the AH mass provides the lower bound on the black hole mass of the final state.  In other words, the quantity $M_{\rm
ADM}-M_{\rm AH}$ is the upper bound on gravitational radiation energy
in the collision process.  Table \ref{lower-bound} shows the value of
$1-M_{\rm AH}/M_{\rm ADM}$ on the AH critical line for
$P/M_0=0,0.5,1.0$.  As the value of $D$ increases, $1-M_{\rm AH}/M_{\rm ADM}$
becomes large for fixed values of $P/M_0=0.5$ and $1.0$ on the AH
critical line.  The area theorem provides a stricter condition for
smaller value of $D$. 

%======================================%
%<<<<<<<<<<<< SECTION III  >>>>>>>>>>>>>>%
%======================================%
\section{Close-slow analysis}

Next we compute gravitational waves in the head-on collision of two
black holes adopting the close-slow approximation. In this
approximation, we assume that $z_0 \ll r_h(M_0)$, $P \ll M_0$, and
$z_0/r_h(M_0) \sim P/M_0$, and evaluate gravitational wave energy up
to order of $(z_0/r_h(M_0))^2$ using a linear perturbative approach.
In the following, the gravitational radius of the system $r_h(M_0)$ is
used as the unit of the length [i.e., $r_h(M_0)=1$] unless specified.

\subsection{Close-slow form of the initial data}

Since we analyze gravitational waves in the Regge-Wheeler type method,
$\widehat{K}_{\mu\nu}^{(\pm)}$ in the spherical-polar coordinate
$(R,\theta,\phi_*), (\phi_*=\phi_1,...,\phi_{n-1})$, should be
derived: 
\begin{multline}
\frac{4M_0}{(n+1)P}\widehat{K}^{R(\pm)}_{R}
=\frac{\mp 2}{R_\pm^{n+1}}\cos\theta(R\mp z_0\cos\theta)\\
+\frac{z_0\mp R\cos\theta}{R_\pm^{n+3}}
\left[
(n-1)(R\mp z_0\cos\theta)^2-R_\pm^2
\right],
\end{multline}
\begin{multline}
\frac{4M_0}{(n+1)P}\frac{\widehat{K}^{(\pm)}_{R\theta}}{R}
=\frac{\pm1}{R_\pm^{n+1}}\sin\theta(R\mp 2z_0\cos\theta)\\
+\frac{z_0\mp R\cos\theta}{R_\pm^{n+3}}
\left[
\pm(n-1)z_0\sin\theta(R\mp z_0\cos\theta)
\right],
\end{multline}
\begin{equation}
\frac{4M_0}{(n+1)P}\widehat{K}^{\theta(\pm)}_{\theta}
=\frac{2}{R_\pm^{n+1}}z_0\sin^2\theta
+\frac{z_0\mp R\cos\theta}{R_\pm^{n+3}}
\left[
(n-1)z_0^2\sin^2\theta-R_\pm^2
\right],
\end{equation}
and
\begin{equation}
\frac{4M_0}{(n+1)P}\widehat{K}^{\phi_*(\pm)}_{\phi_*}
=-\frac{z_0\mp R\cos\theta}{R_\pm^{n+1}}.
\end{equation}
For $z_0 \ll 1$, $K_{\mu\nu}=K_{\mu\nu}^{(+)}+K_{\mu\nu}^{(-)}$
is expanded as 
\begin{equation}
\widehat{K}_{R}^R
=(z_0P/M_0)\frac{n+1}{2}
\left[
n-2-(n^2+n-2)\cos^2\theta
\right]R^{-(n+1)}+O(z_0^2P/M_0),
\label{Krr-close}
\end{equation}
\begin{equation}
\widehat{K}_{\theta}^{\theta}
=(z_0P/M_0)\frac{n+1}{2}
\left[
(n-1)\cos^2\theta+1
\right]R^{-(n+1)}+O(z_0^2P/M_0),
\label{Ktt-close}
\end{equation}
\begin{equation}
\widehat{K}_{\phi_*}^{\phi_*}
=(z_0P/M_0)\frac{n+1}{2}
\left[
(n+1)\cos^2\theta-1
\right]R^{-(n+1)}+O(z_0^2P/M_0),
\label{Kpp-close}
\end{equation}
and $\widehat{K}_{R\theta}=O(z_0^3P/M_0)$. The leading-order term of
$\widehat{K}_{ab}$ is found to be $O(z_0P/M_0)$ and hence 
the right hand side of the Hamiltonian constraint \eqref{Hamiltonian}
is of order $O(z_0^2P^2/M_0^2)$. In the close-slow approximation
adopted here, such terms are higher order and we ignore them. Thus, 
$\psi=0$ and $M_{\rm ADM}= M_0$ in this approximation.

As a result, the conformal factor is given by the Brill-Lindquist one:
\begin{equation}
\varPsi\simeq \varPsi_{\rm
BL}=1+\frac18\left(\frac{1}{R_+^{n-1}}+\frac{1}{R_-^{n-1}}\right).
\end{equation}
By transforming from the isotropic coordinate to the
Schwarzschild-like coordinate
\begin{equation}
r=R\varPsi_0^{2/(n-1)},~~\varPsi_0=1+\frac{1}{4R^{n-1}},
\label{hat-Psi-0}
\end{equation} 
we find that the system is regarded as a perturbed Schwarzschild black
hole 
\begin{equation}
ds^2\simeq\left(\frac{\varPsi_{\rm BL}}{{\varPsi}_0}\right)^{4/(n-1)}
\left[\frac{dr^2}{f(r)}+r^2(d\theta^2+\sin^2\theta d\Omega_{n-1}^2)\right],
~~f(r)=1-\frac{1}{r^{n-1}},
\label{metric-Sch-like}
\end{equation}
\begin{equation}
\left(\frac{\varPsi_{\rm BL}}{{\varPsi}_0}\right)^{4/(n-1)}
= 1+\frac{1/(n-1)R^{n-1}}{1+1/4R^{n-1}}\left(\frac{z_0}{R}\right)^2
C_2^{[(n-1)/2]}(\cos\theta)+O(z_0^4),
\label{metric-Sch-like-2}
\end{equation}  
where $C_\ell^{[\lambda]}$ denotes the Gegenbauer polynomials
defined by the generating function
\begin{equation}
(1-2xt+t^2)^{-\lambda}=\sum_{\ell=0}^{\infty}C_\ell^{[\lambda]}(x)t^\ell.
\end{equation}
Note that the metric~\eqref{metric-Sch-like} with \eqref{metric-Sch-like-2}
is the same as that in our previous analysis of the time-symmetric
initial data~\cite{YSS05}. However, the time-asymmetry is present 
because of the presence of nonzero $\widehat{K}_{\mu\nu}$. 

As found above, the order of the perturbation of the initial metric is
$O(z_0^2)$ and of the extrinsic curvature is $O(z_0P/M_0)$. In the
following, we consider the situation where both $z_0$ and $P/M_0$ 
have the same order.  Under this condition, we can evolve 
the system using a standard perturbation method in the Schwarzschild 
spacetime. 
%This is because a static black hole without angular momentum is 
%uniquely specified by that solution (the uniqueness theorem \cite{Gibbons}). 
From Eqs.~\eqref{Krr-close}--\eqref{Kpp-close}, 
\eqref{metric-Sch-like}, and \eqref{metric-Sch-like-2}, the
leading order of the perturbation contains only the $\ell=2$ mode.

\subsection{Time evolution by the master equation}

The gauge-invariant method for the perturbation around the
Schwarzschild black hole was developed by Kodama and
Ishibashi~\cite{KI03}. They derived a master equation for a variable
$\Phi$, which is related to the gauge-invariant quantities of the
perturbation, as 
\begin{equation}
\frac{\partial^2\Phi}{\partial t^2}-\frac{\partial^2\Phi}{\partial r_*^2}
+V_S\Phi=0, \label{mastereq}
\end{equation}
where
\begin{equation}
V_S(r)=\frac{f(r)Q(r)}{16r^2H^2(r)},
\end{equation}
and 
\begin{equation}
H(r)=m+(1/2)n(n+1)x,~~~x=1/r^{n-1},
\label{def-x}
\end{equation}
\begin{equation}
m=k^2-n,~~~k^2=\ell(\ell+n-1),\label{def-kk}
\end{equation}
\begin{multline}
Q(r)=n^4(n+1)^2x^3
+n(n+1)\left[4(2n^2-3n+4)m+n(n-2)(n-4)(n+1)\right]x^2\\
-12n\left[(n-4)m+n(n+1)(n-2)\right]mx+16m^3+4n(n+2)m^2.
\end{multline}
$r_*$ denotes the tortoise coordinate defined by
\begin{equation}
r_*=\int \frac{dr}{f(r)}.
\end{equation}

%%%%%%%%%%%%%%%%%%%%%%%%%%%%%%%%%%%%%%%%%%%%%%%%%%%%%%%%
\begin{figure}[tb]
\centering
{
\includegraphics[width=0.45\textwidth]{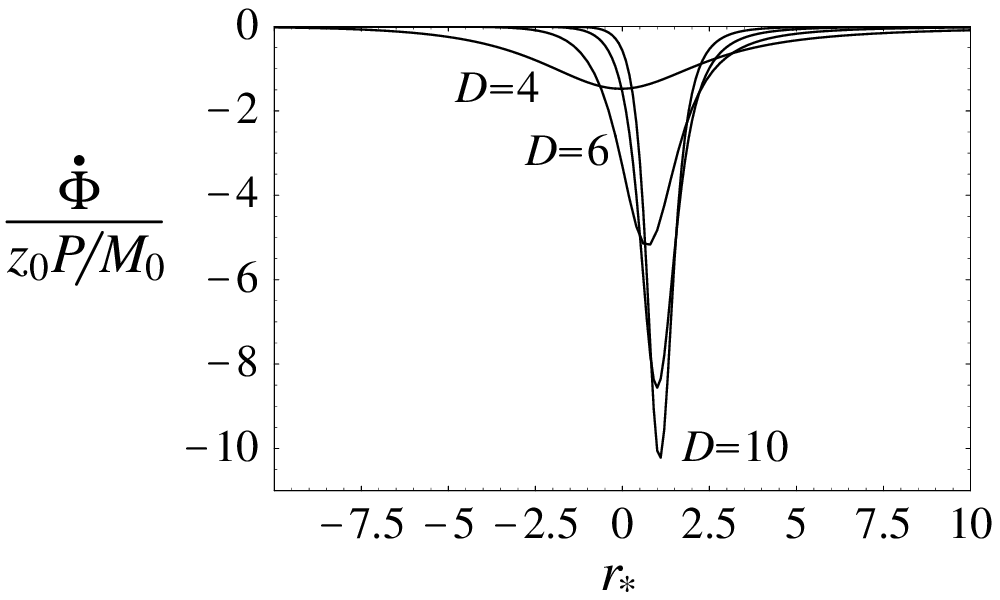}
\includegraphics[width=0.45\textwidth]{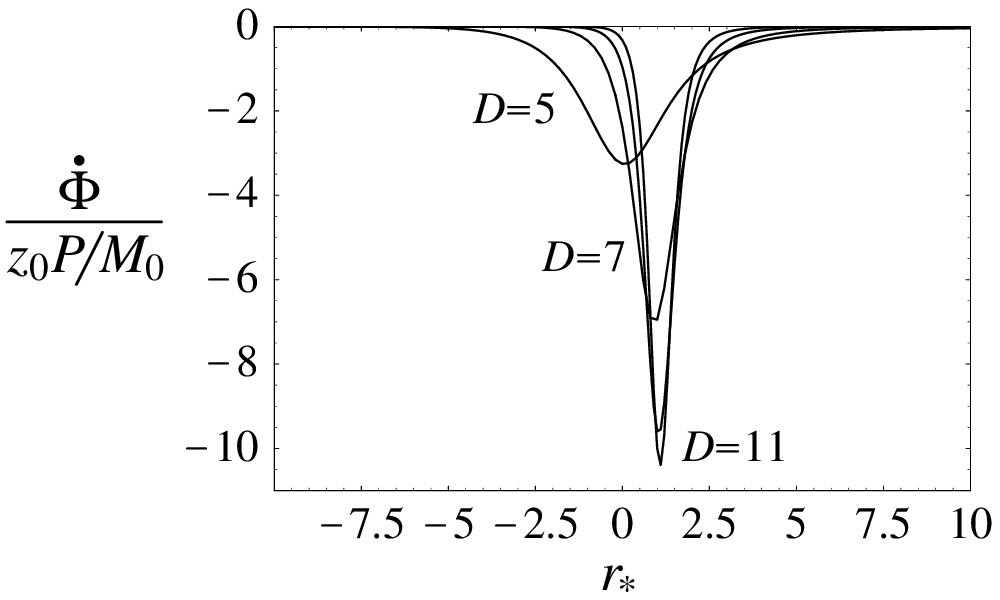}
}
\caption{Time derivative of the master variable $\Phi$ in the unit of $z_0P/M_0$
for $D\equiv n+2=4,6,8,10$ (left) and for $D=5,7,9,11$ (right). For $P>0$, 
$\dot{\Phi}$ is negative.}
\label{initial-BY}
\end{figure}
%%%%%%%%%%%%%%%%%%%%%%%%%%%%%%%%%%%%%%%%%%%%%%%%%%%%%%%%

Initial values of $\Phi$ and $\dot{\Phi}$ (a dot denotes the time
derivative hereafter) are related to the metric perturbation and
$\widehat{K}_{ab}$, respectively. We describe the detail in Appendix
B. The equation for $\Phi(0,r)$ is
the same as that in the Brill-Lindquist case \cite{YSS05}
and the solution is given by
\begin{equation}
{\Phi}(0,r)=\left(z_0^2\right)\frac{n}{4(n^2-1)K_2^{[n]}}
\frac{\sqrt{r}\left[n^2+3n+4+n(n+3)\sqrt{f}\right]}{H(r)R^{(n+3)/2}},
\label{initial-Phi}
\end{equation}
where the definition of $K_2^{[n]}$ is given in Eq.~\eqref{XKLN}.
This has the order of $z_0^2$.  
On the other hand, $\dot{\Phi}(0,r)$ is
proportional to $z_0P/M_0$ and its value is obtained by solving
Eq.~\eqref{eq:initial-master-dot-equation} described in Appendix B. 
The solution is
\begin{equation}
\dot{\Phi}(0,r)=-(z_0P/M_0)
\frac{2n}{(n-1)K_2^{[n]}}\frac{\sqrt{f}}{r^{n/2+1}}
\frac{2(n+2)+(n+1)x}{2(n+2)+n(n+1)x}.
\label{initial-dot-Phi}
\end{equation}
We show the behavior of $\dot{\Phi}(0,r_*)$ in Fig.~\ref{initial-BY}. 

Since Eq. (\ref{mastereq}) is linear, $\Phi$ is naturally decomposed
into two parts 
\begin{equation}
\Phi=(z_0^2)\widehat{\Phi}_{\rm
BL}+\left(z_0P/M_0\right)\widehat{\Phi}_{\rm BY}.
\label{master-variable}
\end{equation}
The solution for $\widehat{\Phi}_{\rm BL}$ is the same as that
in the time-symmetric case derived in our previous paper \cite{YSS05}
\footnote{In \cite{YSS05}, $\Phi(0,r_*)$ was solved numerically. 
We recalculated the temporal evolution using the
analytic formula \eqref{initial-Phi} of the initial condition and
found that the difference between the two is $\sim 10^{-5}\%$.}. 
Here, we show only the computation of $\widehat{\Phi}_{\rm BY}$. 
For the numerical computation,
we use the second-order finite differencing code developed in
\cite{YSS05} and solve the equation in the domain $-200\le r_*\le
1000$ with the grid spacings $dr_*=0.01$ ($D=4$--$7$) and $0.005$
($D=8$--$11$) and $dt=0.2dr_*$. Computation was performed changing the
grid spacing and we confirmed that the numerical results converge at
second order. For the chosen grid spacing, the error 
evaluated with $\epsilon_1$ in Eq.~\eqref{error-estimate} is $0.05\%$--$0.9\%$
for $D=4$--$7$ and $0.4\%$--$2\%$ for $D=8$--$11$. The error in the values
listed in Table II is $\lesssim 0.1\%$.

\subsection{Numerical results}

%%%%%%%%%%%%%%%%%%%%%%%%%%%%%%%%%%%%%%%%%%%%%%%%%%%%%%%%
\begin{figure}[tb]
\centering
{
\includegraphics[width=0.4\textwidth]{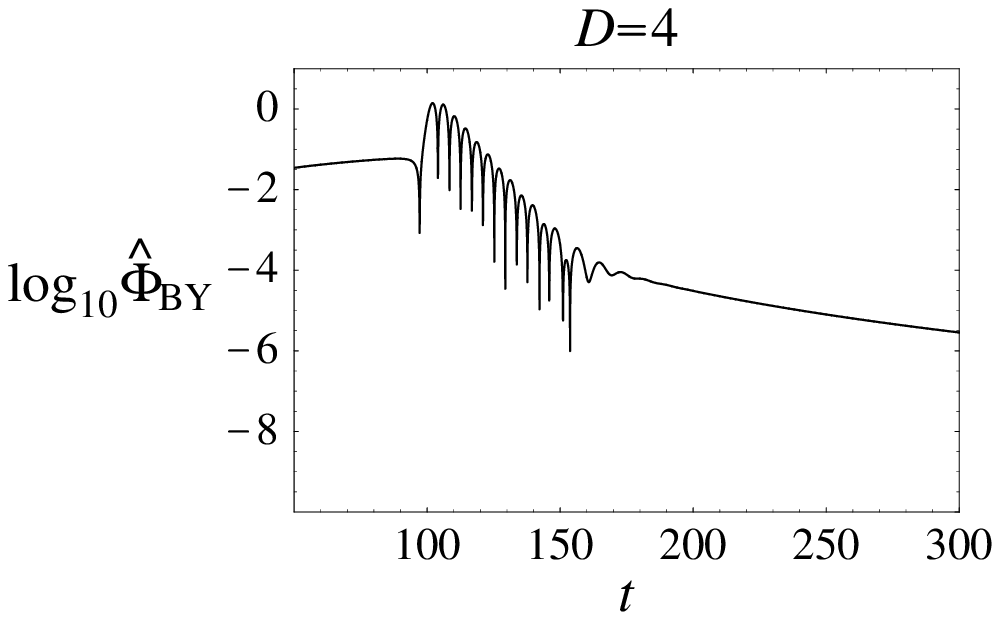}\hspace{5mm}
\includegraphics[width=0.4\textwidth]{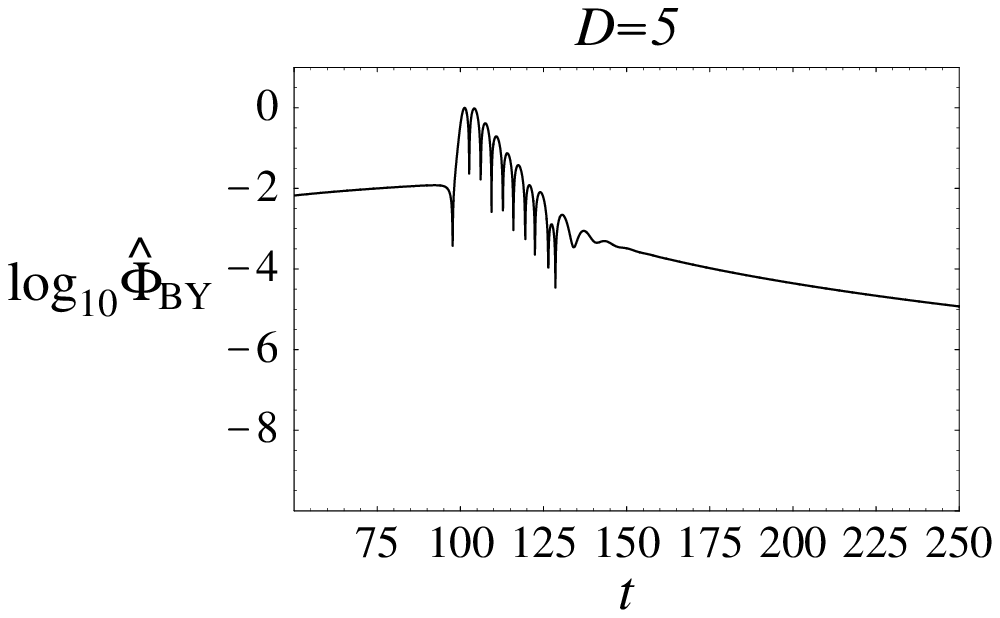}
\includegraphics[width=0.4\textwidth]{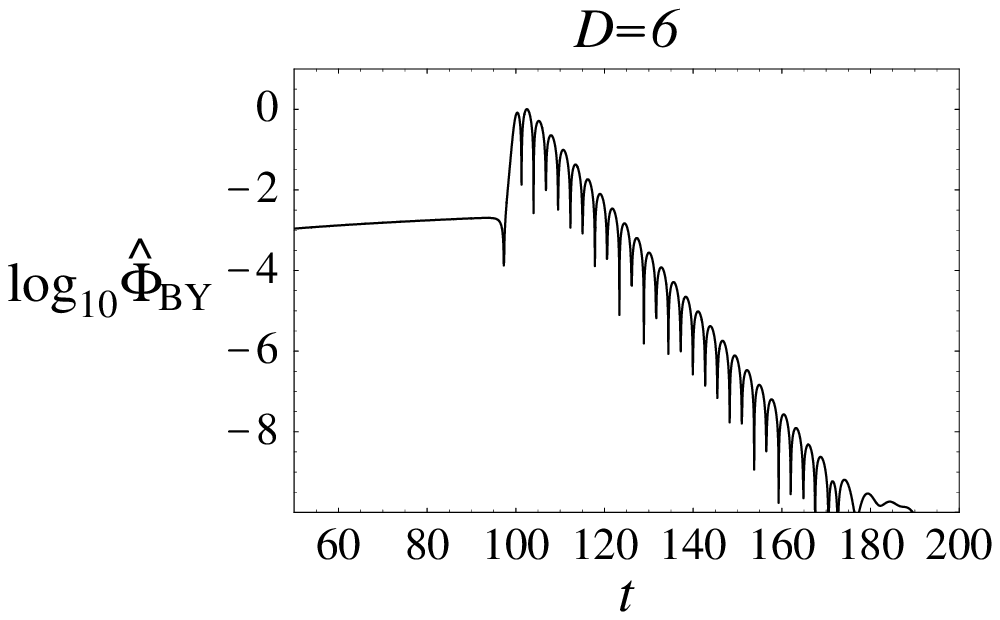}\hspace{5mm}
\includegraphics[width=0.4\textwidth]{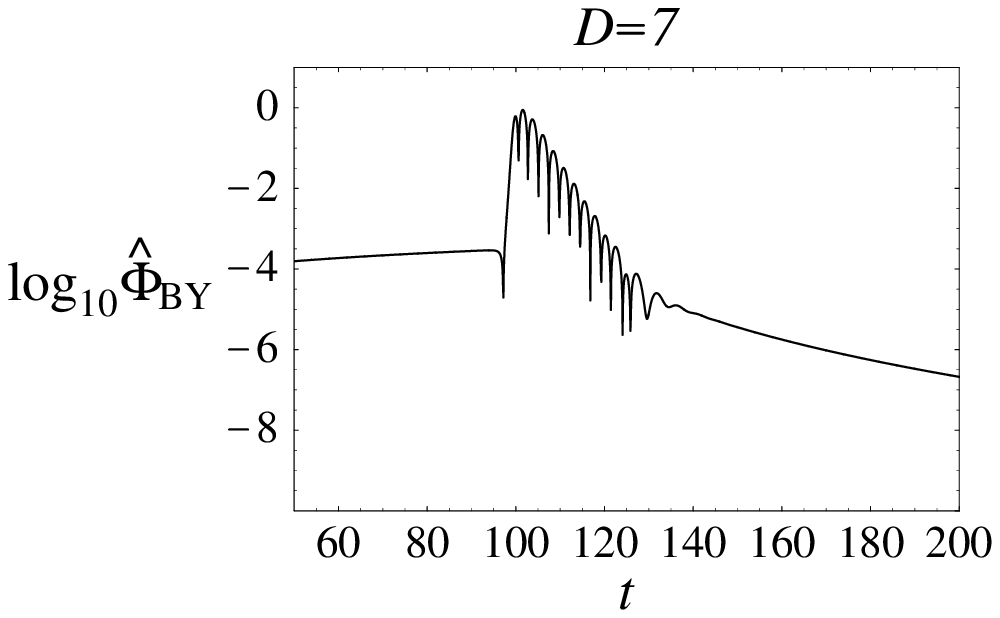}
\includegraphics[width=0.4\textwidth]{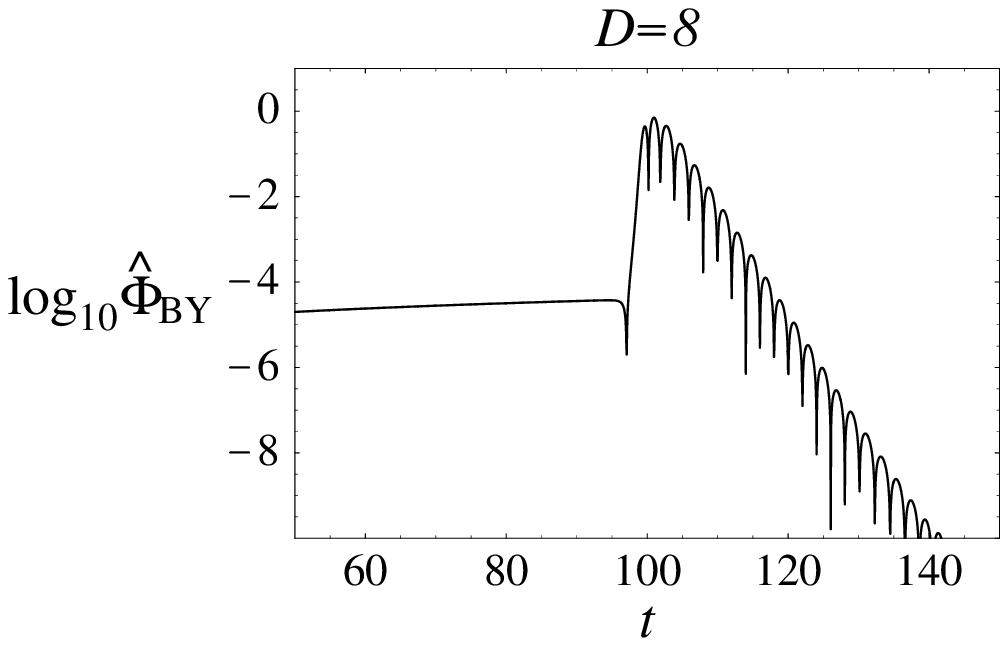}\hspace{5mm}
\includegraphics[width=0.4\textwidth]{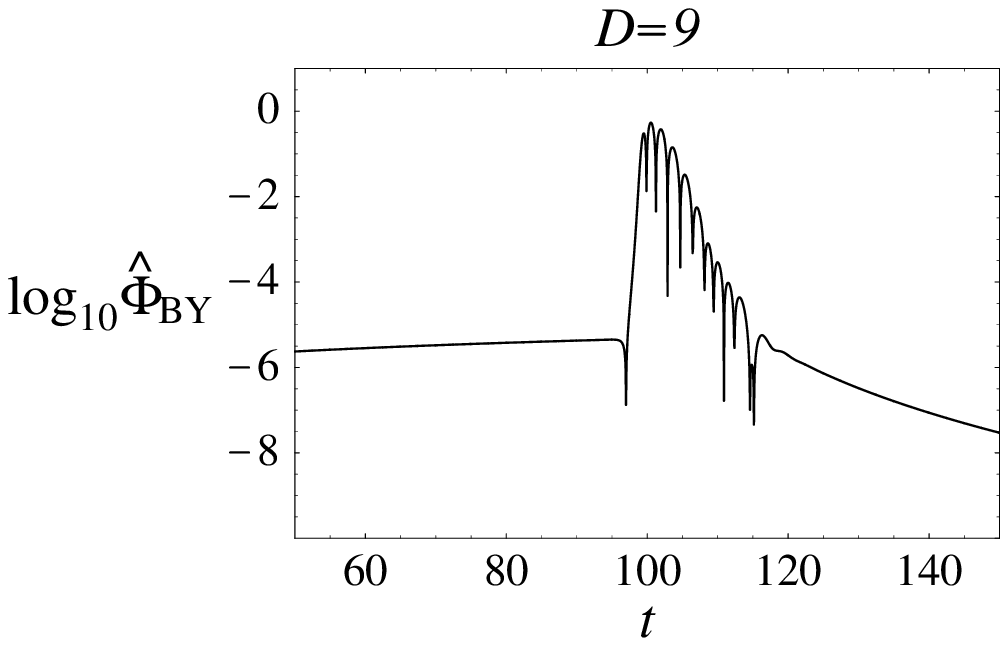}
\includegraphics[width=0.4\textwidth]{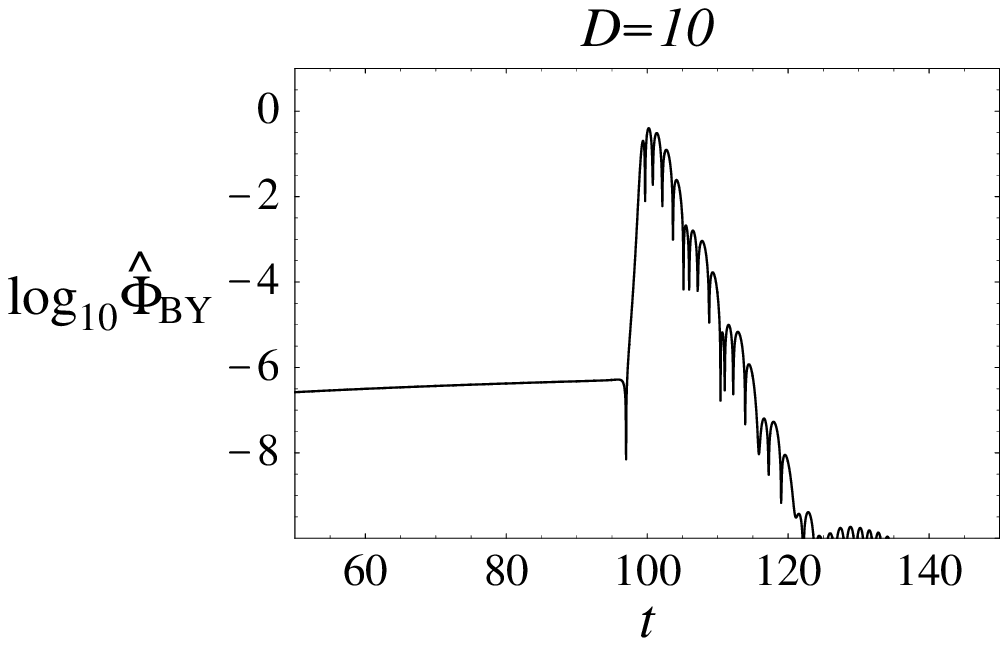}\hspace{5mm}
\includegraphics[width=0.4\textwidth]{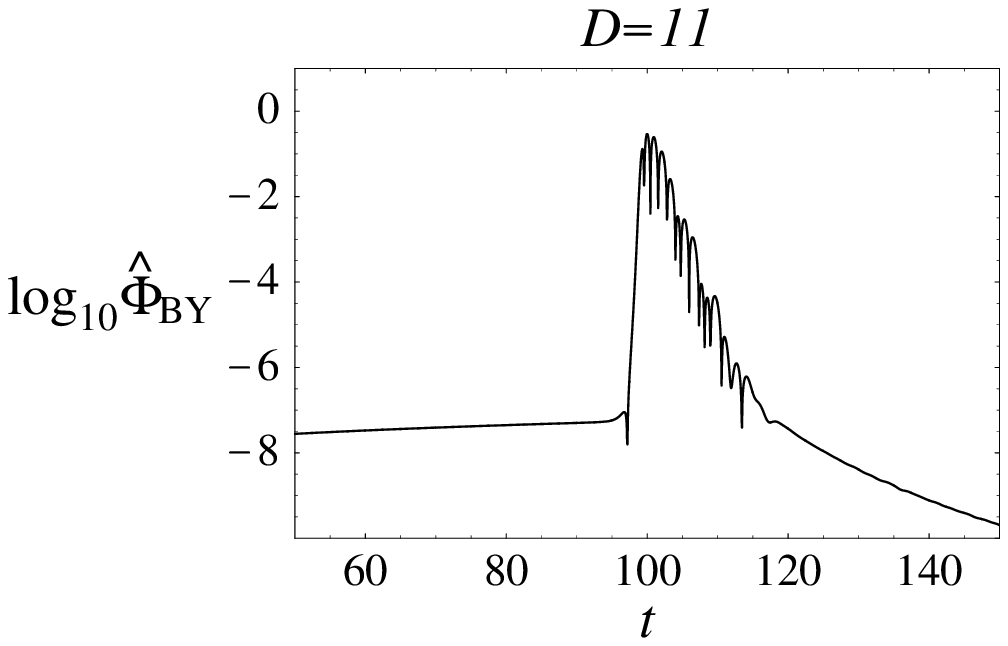}
}
\caption{The time evolution of $\widehat{\Phi}_{\rm BY}$
for $D=4$--$11$ observed at $r_*=100$. }
\label{amp-BY}
\end{figure}
%%%%%%%%%%%%%%%%%%%%%%%%%%%%%%%%%%%%%%%%%%%%%%%%%%%%%%%%

%%%%%%%%%%%%%%%%%%%%%%%%%%%%%%%%%%%%%%%%%%%%%%%%%%%%%%%%
\begin{figure}[tb]
\centering
{
\includegraphics[width=0.45\textwidth]{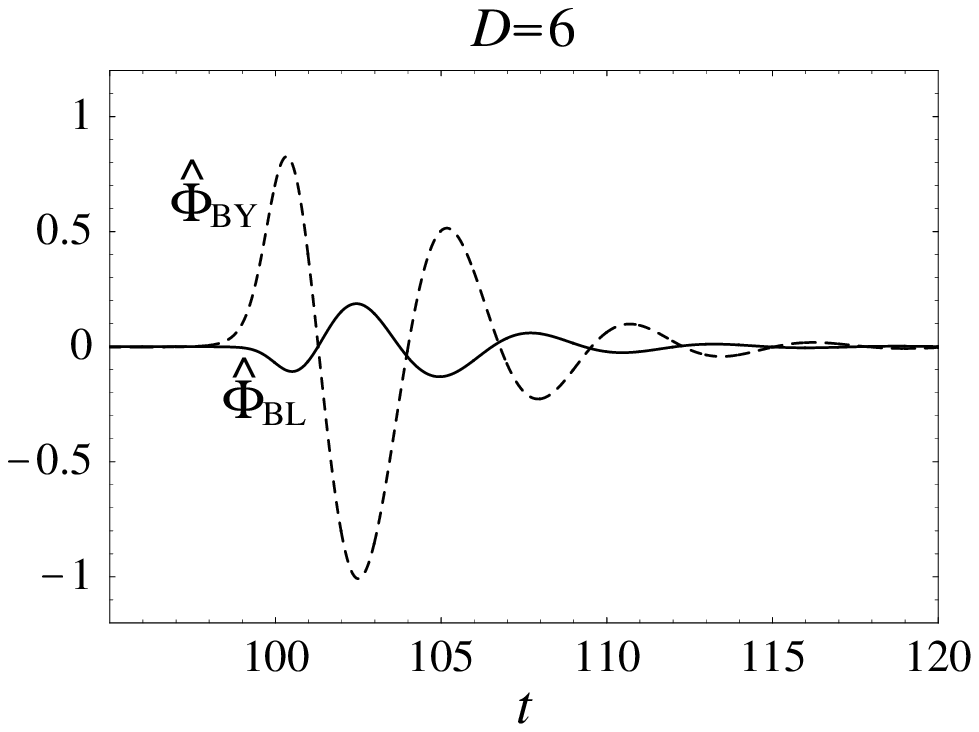}
\includegraphics[width=0.45\textwidth]{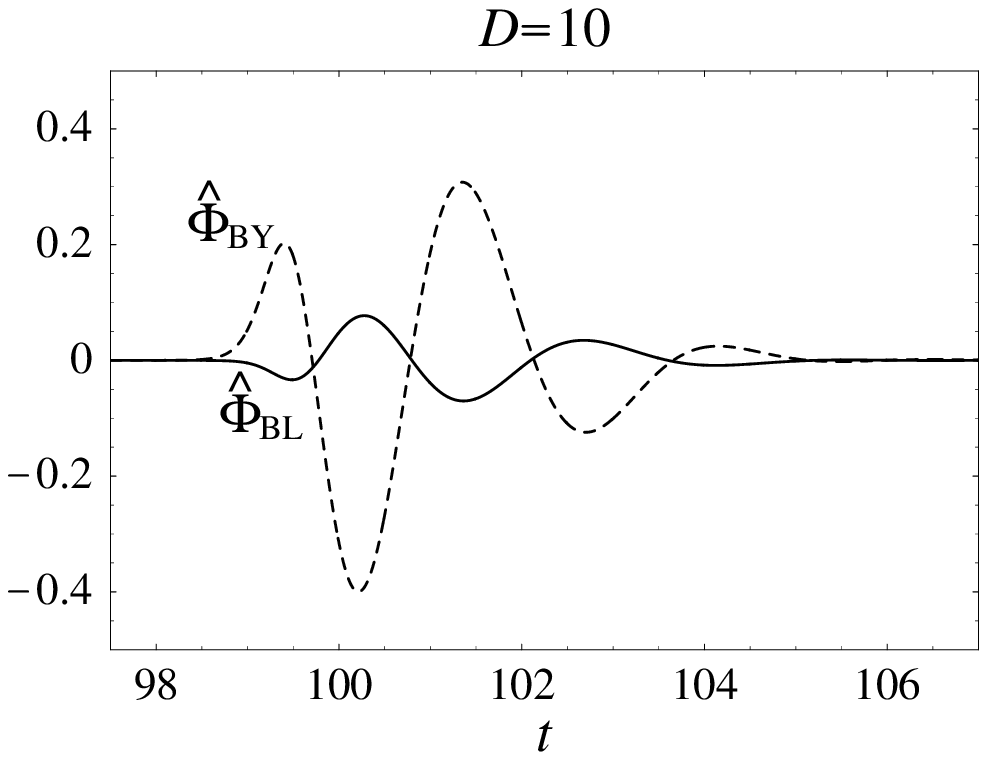}
}
\caption{The time evolution of $\widehat{\Phi}_{\rm BY}$ (dashed line) and 
$\widehat{\Phi}_{\rm BL}$ (solid line) for $D=6$ (left) and $10$ (right)
observed at $r_*=100$.}
\label{compare-BLBY}
\end{figure}
%%%%%%%%%%%%%%%%%%%%%%%%%%%%%%%%%%%%%%%%%%%%%%%%%%%%%%%%

Figure~\ref{amp-BY} shows numerical results for the time evolution
of $\widehat{\Phi}_{\rm BY}$. Soon after the onset of the calculation, 
a quasi-normal mode is excited irrespective of dimension and,
subsequently, the power-law tail is seen for four and odd dimensions.
We read off the quasinormal frequencies $\omega_{\rm QN}$ and checked
the consistency with previous results of $\omega_{\rm QN}$ listed
in~\cite{YSS05, Konoplya, BCG04}.  The behaviors of $\widehat{\Phi}_{\rm BL}$ and
$\widehat{\Phi}_{\rm BY}$ are compared in Fig.~\ref{compare-BLBY} for
$D=6$ and $10$.  It is found that phases of $\widehat{\Phi}_{\rm BL}$
and $\widehat{\Phi}_{\rm BY}$ disagree and the phase shift is $\approx
\pi$. This implies that two terms interfere each other. This feature
universally holds irrespective of the dimensionality. 

%%%%%%%%%%%%%%%%%%%%%%%%%%%%%%%%%
\begin{table}[tb]
\centering
\caption{The values of $c_1$, $c_2$ and $c_3$ of Eq.~\eqref{Erad-c1c2c3} 
for $D=4$--$11$. }
\begin{ruledtabular}
\begin{tabular}{c|cccccccc}
$D$ & $4$ & $5$ & $6$ & $7$ & $8$ & $9$ & $10$ & $11$  \\
  \hline 
$c_1$ & $0.0252$ & $0.0245$ & $0.0290$ & $0.0288$ & $0.0258$ & $0.0223$ & $0.0194$ & $0.0172$ \\
$c_2$ & $-0.165$ & $-0.243$ & $-0.294$ & $-0.287$ & $-0.251$ & $-0.213$ & $-0.182$ & $-0.158$ \\
$c_3$ & $0.343$ & $0.671$ & $0.808$ & $0.765$ & $0.647$ & $0.539$ & $0.456$ & $0.396$ 
  \end{tabular}
    \end{ruledtabular}
  \label{c1c2c3}
\end{table}
%%%%%%%%%%%%%%%%%%%%%%%%%%%%%%%%%

From $\widehat{\Phi}_{\rm BL}$ and $\widehat{\Phi}_{\rm BY}$, we
calculate the radiated energy of gravitational waves by the following
formula (see \cite{YSS05, BCG04} for a derivation):
\begin{equation}
E_{\rm rad}=\frac{k^2(n-1)(k^2-n)}{32\pi nG}\int\dot{\Phi}^2dt.
\end{equation}
Substituting Eq. \eqref{master-variable} into the above formula,
$E_{\rm rad}$ is rewritten as
\begin{equation}
\frac{E_{\rm rad}}{M_0}=c_1z_0^4+c_2z_0^3(P/M_0)
+c_3z_0^2(P/M_0)^2,
\label{Erad-c1c2c3}
\end{equation}
where $c_1$, $c_2$, and $c_3$ are constants determined by numerical
integration.  These values are listed in Table~\ref{c1c2c3}. The
formula \eqref{Erad-c1c2c3} together with Table~\ref{c1c2c3} will be
used for the benchmark of the fully nonlinear analysis in numerical
relativity as in the four-dimensional case~\cite{PP94, BAABPPS97}.

%%%%%%%%%%%%%%%%%%%%%%%%%%%%%%%%%%%%%%%%%%%%%%%%%%%%%%%%
\begin{figure}[tb]
\centering
{
\includegraphics[width=0.33\textwidth]{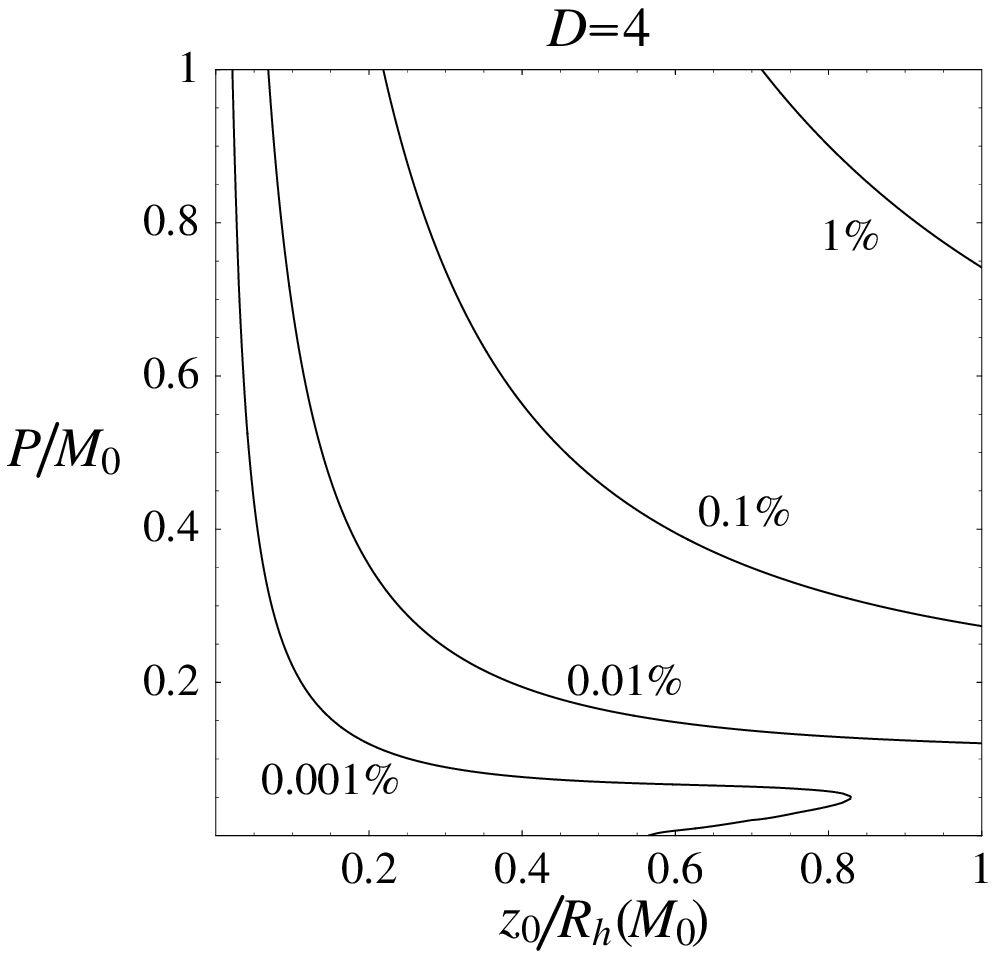}\hspace{5mm}
\includegraphics[width=0.33\textwidth]{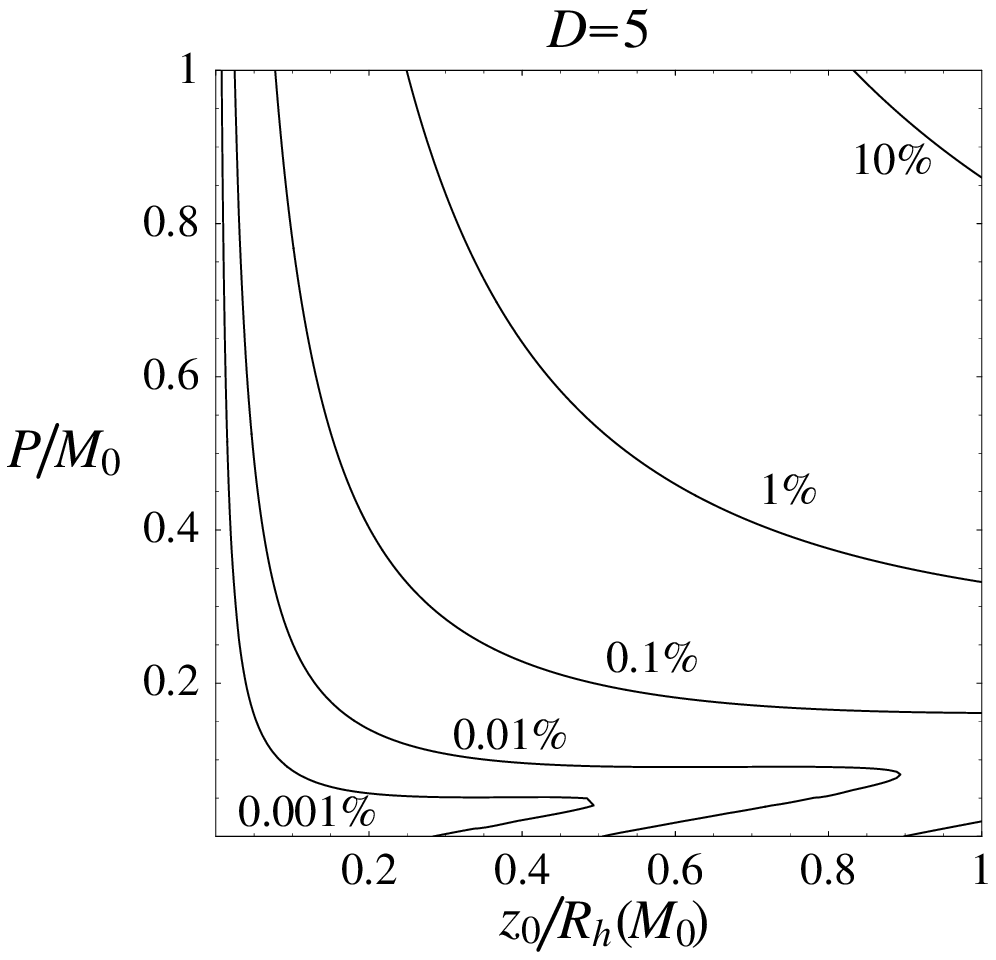}\\
\includegraphics[width=0.33\textwidth]{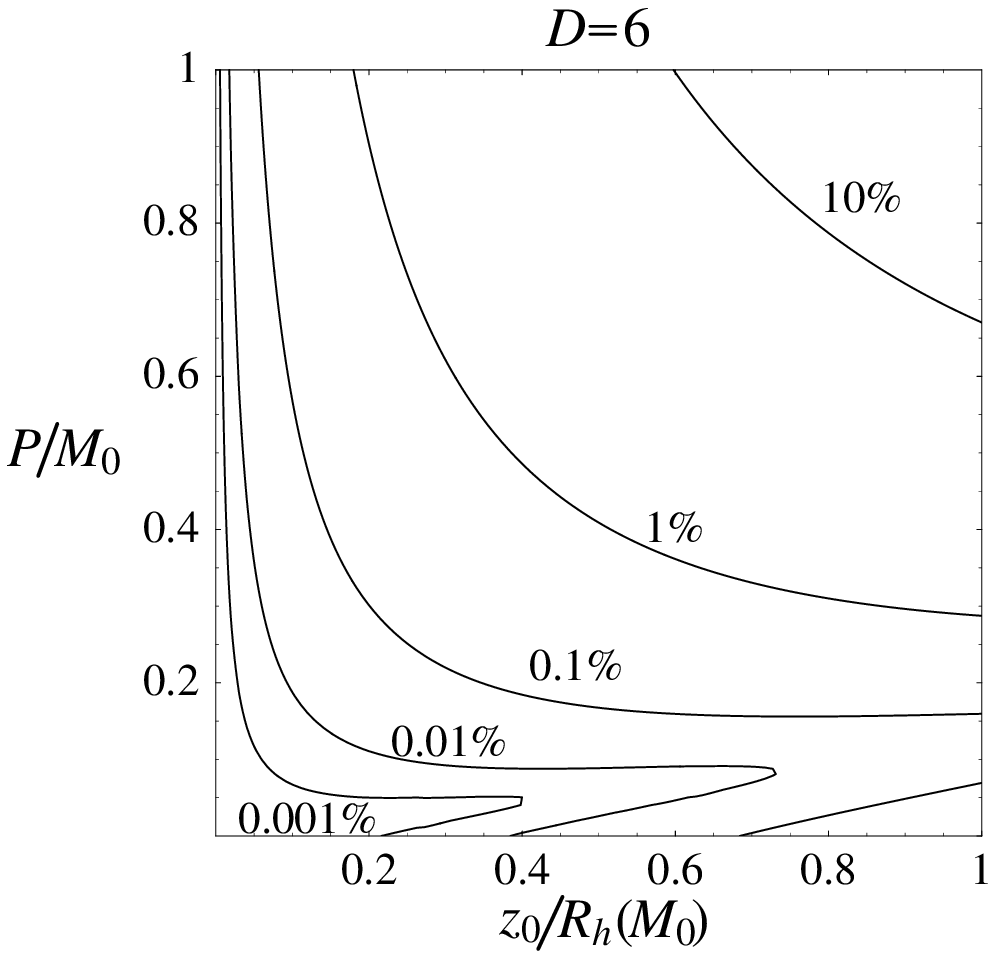}\hspace{5mm}
\includegraphics[width=0.33\textwidth]{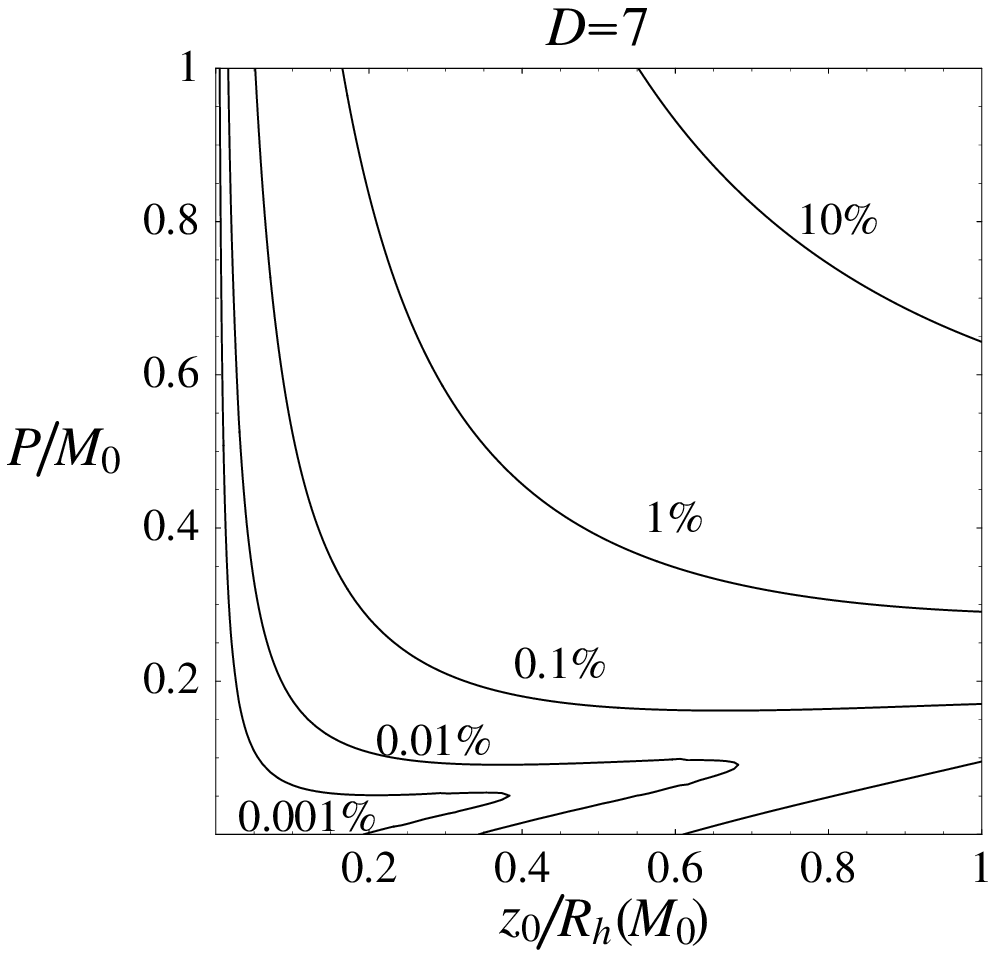}\\
\includegraphics[width=0.33\textwidth]{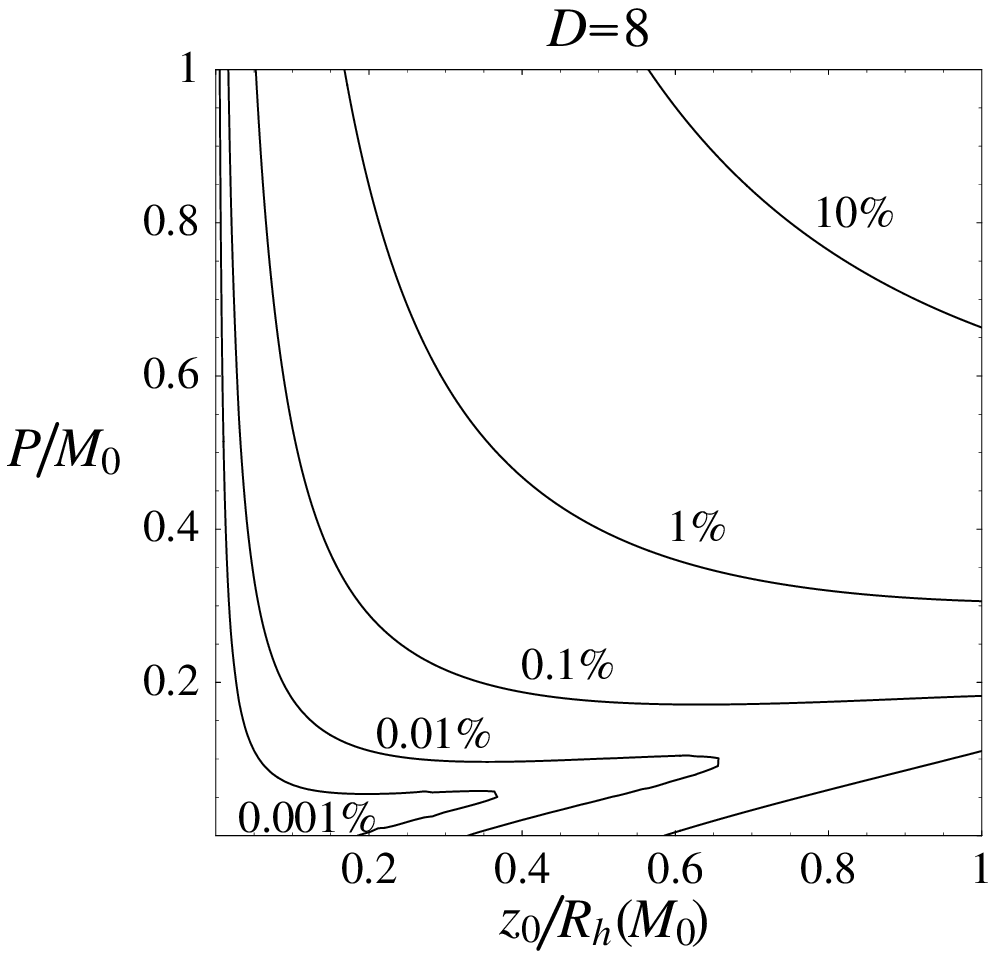}\hspace{5mm}
\includegraphics[width=0.33\textwidth]{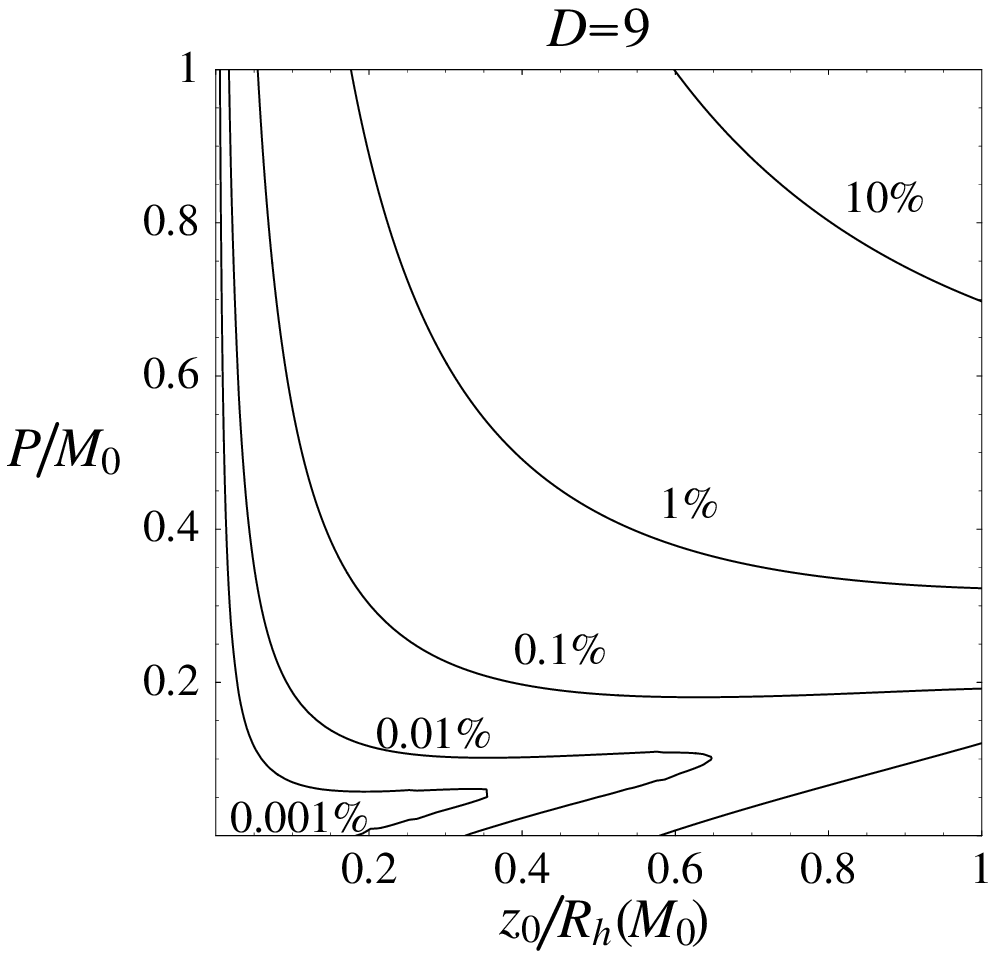}\\
\includegraphics[width=0.33\textwidth]{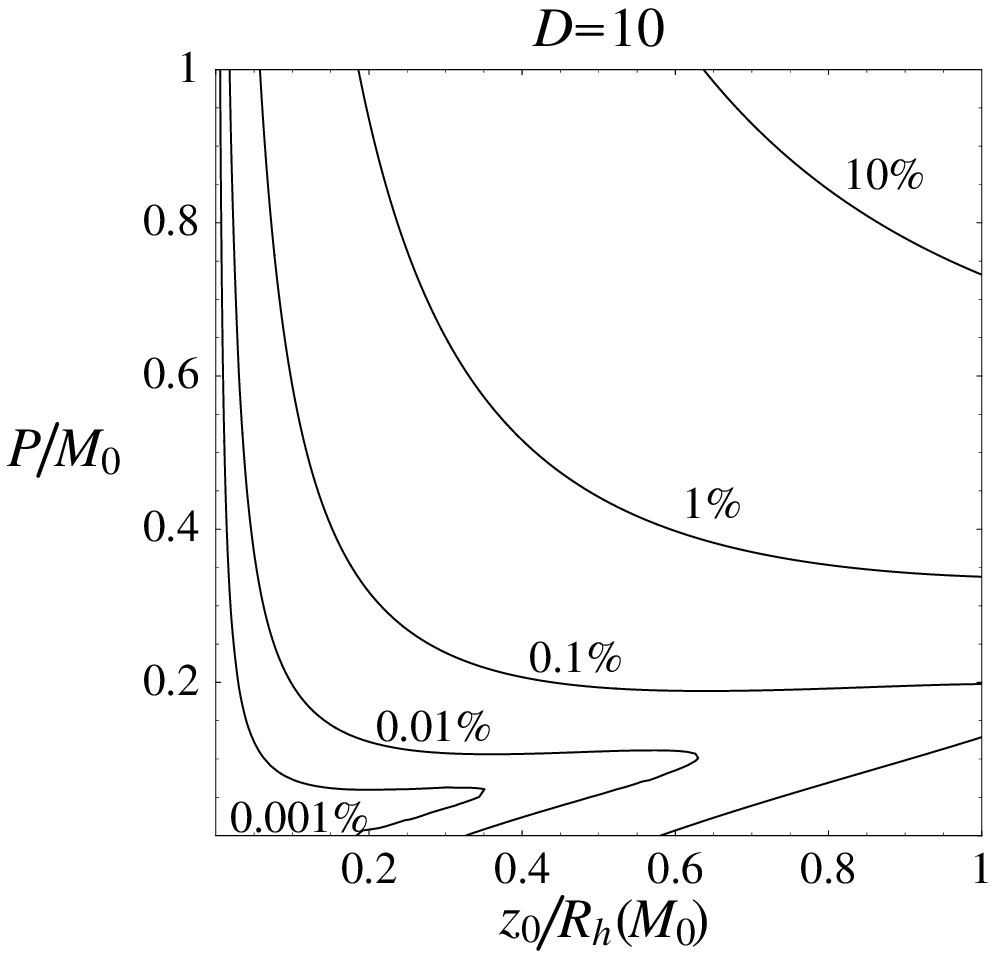}\hspace{5mm}
\includegraphics[width=0.33\textwidth]{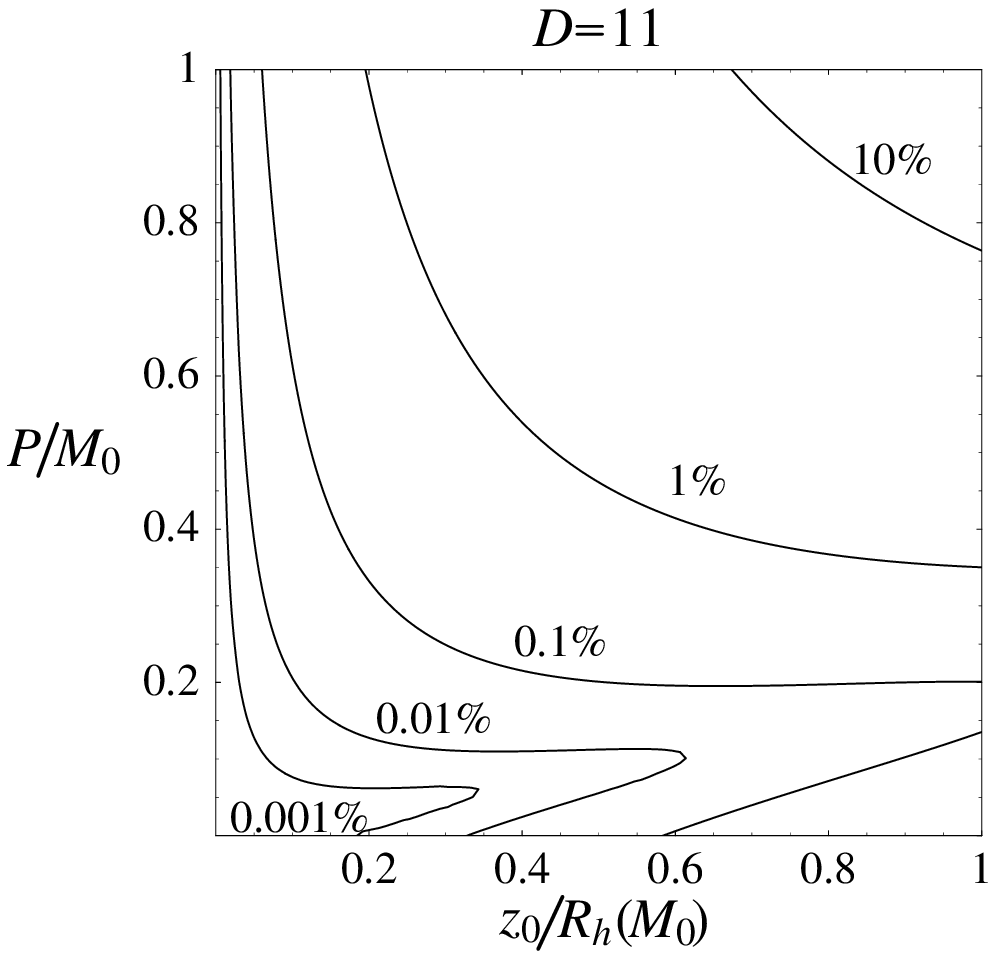}
}
\caption{Contours of radiation efficiency $E_{\rm rad}/M_0$ (shown in the unit of
\%) on the $(z_0/R_h(M_0), P/M_0)$-plane for $D=4$--$11$
predicted by the close-slow analysis. 
Note that the unit of $z_0$ is 
$R_h(M_0)$ in this figure.}
\label{efficiency-zP-plane}
\end{figure}
%%%%%%%%%%%%%%%%%%%%%%%%%%%%%%%%%%%%%%%%%%%%%%%%%%%%%%%%

Figure~\ref{efficiency-zP-plane} shows the contours of $E_{\rm
rad}/M_0$ on the $(z_0/R_h(M_0),P/M_0)$-plane. For a fixed value of
$z_0$, the radiated energy decreases as the value of $P/M_0$ is increased
for small values of $P$ as $P/M_0 \lesssim 0.2\times z_0$. This is
because the phases of $\widehat{\Phi}_{\rm BL}$ and
$\widehat{\Phi}_{\rm BY}$ disagree by a factor of $\approx \pi$
(cf. Fig. 6).  In the range $P/M_0 \gtrsim 0.2\times z_0$, the
amplitude of second term in Eq.~\eqref{master-variable} exceeds the
first term and $E_{\rm rad}$ increases as $P/M_0$ is increased.  The
similar behavior was reported in the four dimension case
\cite{BAABPPS97}. 

\subsection{Dependence on dimensionality}

To get some insight for the dependence of radiation efficiency on the
value of $D$, we evaluate $E_{\rm rad}/M_0$ by choosing characteristic
values of $z_0$ and $P$ for $D=4$--$11$. In comparison, we fix the
value of $P/M_0$ since $P/2M_0$ could be interpreted as the value
of momentum divided by the rest mass, 
i.e., $v/\sqrt{1-v^2}$ 
where $v$ is the velocity of each incoming black hole. We adopt $P/M_0=0, 0.5$
and $1$ which corresponds to $\gamma=1$, $\sqrt{2}\simeq 1.41$, and 
$\sqrt{5}\simeq 2.24$. Then we
recall Fig.~\ref{critical-AH} and evaluate $E_{\rm rad}$ on the AH
critical line, since the close-slow approximation holds for the system
sufficiently close to the Schwarzschild spacetime and it is necessary
to choose $z_0$ for which the common AH presents.

%%%%%%%%%%%%%%%%%%%%%%%%%%%%%%%%%
\begin{table}[tb]
\centering
\caption{The values of $E_{\rm rad}/M_0$ evaluated on the AH critical line
for $P/M_0=0.0,0.5,1.0$. The unit is \%. }
\begin{ruledtabular}
\begin{tabular}{c|cccccccc}
$D$ & $4$ & $5$ & $6$ & $7$ & $8$ & $9$ & $10$ & $11$  \\
  \hline 
$P/M_0=0.0$ & $0.0034$ & $0.059$ & $0.20$ & $0.34$ & $0.44$ & $0.49$ & $0.51$ & $0.52$ \\
$P/M_0=0.5$ & $0.30$ & $1.76$ & $2.83$ & $3.01$ & $2.80$ & $2.53$ & $2.28$ & $2.08$ \\
$P/M_0=1.0$ & $1.6$ & $7.3$ & $11.5$ & $12.2$ & $12.0$ & $11.4$ & $10.5$ & $9.8$ \\
  \end{tabular}
  \end{ruledtabular}
  \label{compare-Erad}
\end{table}
%%%%%%%%%%%%%%%%%%%%%%%%%%%%%%%%%

Table~\ref{compare-Erad} shows the values of $E_{\rm rad}/M_0$
evaluated in this procedure. In the higher-dimensional cases, it is
$\sim 2\%$--3\% for $P/M_0=0.5$ and $\sim 10\%$ for $P/M_0=1.0$. These
values do not depend significantly on $D$ in the higher-dimensional
cases. Note that the values in Table~\ref{compare-Erad} does not
contradict the bounds derived from the AH area shown in
Table~\ref{lower-bound}.

The difference between $M_0$ and $M_{\rm ADM}$ increases with increase
of $P/M_0$ (cf. Fig.~\ref{contour-ADM}).  This difference comes from
time asymmetry of the initial data and has the magnitude of order
$(z_0P/M_0)^2$. Thus for a large difference with $M_{\rm ADM}/M_0-1 \gtrsim
0.1$, the linear approximation breaks down. In the four-dimensional
case, we know that the nonlinear effect suppresses $E_{\rm
rad}/M_{\rm ADM}$~\cite{BAABPPS97}.  If this is also the case for
$D\ge 5$, the value of $E_{\rm rad}/M_{\rm ADM} \sim 10\%$ for $P/M_0$
would be an overestimate and would indicate that gravitational
radiation do not significantly contribute to the loss of the system energy 
in the head-on collision of two black holes.

%======================================%
%<<<<<<<<<<<< SECTION V  >>>>>>>>>>>>>>%
%======================================%

\section{Discussion}

In this paper, we presented a generalized Bowen-York formulation for
the initial value problem of multi--black holes in higher-dimensional
space.  Using this formulation, we derived analytic solutions of the
momentum constraint equation for extrinsic curvature of moving black
holes in the conformally flat higher-dimensional space. As an
application, initial data for head-on collision of two equal-mass
black holes were computed numerically solving the Hamiltonian constraint
for the conformal factor.  The properties of the obtained space such
as the ADM mass and the common AH were analyzed. We determined the
critical line for formation of a common AH in the parameter space
of $(z_0,P)$. 

Then, we evolved this system using the close-slow approximation and
analyzed gravitational waves. We derived the formula for the radiated
energy for a wide variety of $D$ [see Eq.~\eqref{Erad-c1c2c3} together
with Table~\ref{c1c2c3}] and clarified that it depends weakly on $D$;
irrespective of $D$ the efficiency is less than $\sim 10\%$ for
$P/M_0 \leq 1$ and $z_0 \leq r_h(M_0)$.

As we mentioned in Sec. I, one of the motivations of this study is to
clarify the total radiated energy of gravitational waves in the black
hole formation through particle collisions in accelerators.  However,
there are two shortcomings in this study. One is that we use the
linear perturbation method which provides a reliable result only for
small values of $z_0$ and $P$. The other is that the Bowen-York black
holes system does not accurately model the high-energy particles
system.

As for the first problem, we can learn from the four-dimensional
study.  In \cite{BAABPPS97}, the results of the close-slow
approximation and of the fully nonlinear numerical simulation are
compared. It shows that the results in the close-slow approximation
agrees approximately with the fully nonlinear results for small
separation and small momentum. However, the difference becomes larger
as separation or momentum is increased. For $P \gg M_0$, the nonlinear
numerical results indicate that the radiation efficiency increases as
$P$ is increased and approaches to an asymptotic value, while extrapolation
of the linear perturbation results gives infinity. To know the results
for $P \gg M_0$ in higher-dimensional spacetime, obviously, fully
nonlinear numerical simulation is necessary. 

The difference between the high-energy particles and the Bowen-York
black holes is the other problem. As the model of a high-energy
particle, the metric of Aichelburg-Sexl~\cite{AS71} is often used. It
is obtained by boosting the Schwarzschild black hole to the speed of
light with fixed energy $p=\gamma m$ and taking lightlike limit
$\gamma\to \infty$.  The gravitational field is localized in the
transverse plane to the direction of motion and forms a gravitational
shock wave, which is a reminiscent of the infinite Lorentz contraction
of the isotropic gravitational field of the original Schwarzschild
black hole.  The spacetime is flat except at the shock wave and this
enables us to write down the metric of two high-energy particles
outside of the lightcone of shock collision. Hence, the AH formation
at the instant of collision has been investigated so
far~\cite{EG02,YN03,YR05}.

Obviously, gravitational field (the conformal factor or the extrinsic
curvature) is not localized in the case of Bowen-York black holes.
This indicates that the Bowen-York black holes system cannot be a
precise model for the high-energy particles system even in the case
$z_0\gg R_h(M_0)$ and $P\gg M_0$.  Actually, there exist several
discussions~\cite{JDKN03} about the fact that the Bowen-York black
hole system contains unphysical gravitational waves, i.e., so-called
junk energy. Thus, in the fully nonlinear numerical calculation with
the Bowen-York initial data, one should take into account the
possibility that junk energy changes the estimate of radiation
efficiency. To avoid this, one should make the junk energy
radiate away before the collision setting the initial separation
sufficiently large.

Finally, we comment on the importance of the study of the grazing
collision.  In the head-on collision case, the upper bound on the
radiation efficiency ($\lesssim 40\%$) in the head-on collision of
Aichelburg-Sexl particles was obtained~\cite{EG02} by studying the AH
area and applying the area theorem. Although the same discussion was
done in the grazing collision with nonzero impact parameter
$b$~\cite{YN03, YR05}, the upper bound on the radiation efficiency is
very weak for large $b$. Hence, the direct calculation of
gravitational radiation in the grazing collision is more important
than the head-on collision case.  Also, it is required to calculate
the Kerr parameter of the resultant black hole.  In order to solve
these problems, the off-axis collision of the Bowen-York black holes
could be an approximation and provide several implications.  We plan
to study this process using the close-slow approximation as a first
step.

\acknowledgments

The work of H. Y. was partially supported by a Grant for The 21st Century 
COE Program (Holistic Research and Education Center for Physics 
Self-Organization Systems) at Waseda University. H. Y. also thanks the
Killam Trust for the financial support.  
The work of T. S. was supported by Grant-in-Aid for Scientific 
Research from Ministry of Education, Science, Sports and Culture of 
Japan (No. 13135208, No. 14102004, No. 17740136 and No. 17340075), 
the Japan-U.K. and Japan-France Research  Cooperative Program. 

\appendix

\section{Other solutions of extrinsic curvature}

We briefly comment on other solutions of $\widehat{K}_{\mu\nu}$
for the momentum constraint~\eqref{momentum}, which were not
introduced in Sec. II. Since the equations for $B_\mu$ and $\chi$ are
pure Laplace equations, we can make infinite number of solutions for
the extrinsic curvature. Here, we present solution of low-multipole
momenta. 

To make a spinning black hole, we write 
\begin{equation}
B_\mu=\frac{(n-1)\pi G J_{\mu\nu} n^\nu}{n\Omega_nR^n},~~~~\chi=0
\end{equation}
where $J_{\mu\nu}$ is an anti-symmetric tensor. Then, 
$\widehat{K}_{\mu\nu}$ is written as 
\begin{equation}
\widehat{K}_{\mu\nu} =-\frac{4\pi (n+1)G}{\Omega_n R^{n+1}}
\left(J_{\mu\rho}n^\rho n_\nu+ J_{\nu\rho}n^\rho n_\mu\right).
\end{equation}
$J_{\mu\nu}$ denotes an angular momentum tensor expressed by 
\begin{equation}
J_{\mu\nu}=\frac{1}{8\pi G}\int_{R\to \infty}
\left(x_\mu K_{\nu\rho}-x_\nu K_{\mu\rho}\right)n^\rho dS. 
\end{equation}
In the three-dimensional case, we can define the angular momentum vector 
by the formula
\begin{equation}
J_\mu=\frac12\epsilon_{\mu\nu\rho}J^{\nu\rho}.
\end{equation}
The conformal factor $\varPsi$ can be determined by solving the
Hamiltonian constraint in the same procedure as in Sec. III. 

It is possible to derive other solutions, e.g., by setting
\begin{equation}
B_\mu=\frac{C_{\mu\nu}n^\nu}{R^{n}},~~\chi=0,
\end{equation}
or 
\begin{equation}
B_\mu=\frac{C_{\mu\nu\rho}}{R^{n+1}}\left[n_\nu n_\rho
-\frac{\delta_{\nu\rho}}{(n+1)}\right],~~\chi=0,
\end{equation}
where $C_{\mu\nu}$ and $C_{\mu\nu\rho}$ are 
symmetric and anti-symmetric tensors, respectively.
However the physical meaning of these solutions is unclear.

\section{Initial condition of the master variable}

In this appendix, we explain how to set the initial condition of the
master variable $\Phi$.  We begin by studying the initial condition of
the following perturbative metric
\begin{equation}
\delta h_{\mu\nu}=
\left(
\begin{array}{cccc}
fH_0{\mathbb S} & H_1{\mathbb S} & h_0{\mathbb S}_\theta & h_0{\mathbb S}_{\phi_*}\\
{\rm sym.} & f^{-1}H_2{\mathbb S} & h_1 {\mathbb S}_\theta & h_1{\mathbb S}_{\phi_*}\\
{\rm sym.} & {\rm sym.} & 2r^2(H_L\gamma_{\theta\theta}{\mathbb S}+H_T{\mathbb S}_{\theta\theta}) 
&2r^2(H_L\gamma_{\theta\phi_*}{\mathbb S}+H_T{\mathbb S}_{\theta\phi_*})\\
{\rm sym.} & {\rm sym.}&{\rm sym.} &2r^2(H_L\gamma_{\phi_*\phi_*}{\mathbb S}+H_T{\mathbb 
S}_{\phi_*\phi_*}) 
\end{array}
\right)
\label{eq:scalar-perturbation}
\end{equation}
where ${\mathbb S}$ denotes the hyper-spherical harmonics
\begin{equation}
{\mathbb S}=S^{[n]}_\ell
\equiv K^{[n]}_\ell C^{[(n-1)/2]}_\ell(\cos\theta),
\end{equation}
\begin{equation}
K^{[n]}_{\ell}=\left[\frac{4\pi^{(n+1)/2}\Gamma(n+\ell-1)}
{(n+2\ell-1)\Gamma(\ell+1)\Gamma((n-1)/2)\Gamma(n-1)}\right]^{-1/2}, 
\label{XKLN}
\end{equation}
on the unit sphere of which metric is
\begin{equation}
\gamma_{ij}dz^idz^j=d\theta^2+\sin^2\theta d\Omega_{n-1}^2.
\end{equation} 
${\mathbb S}_i$ and ${\mathbb S}_{ij}$ are given by
\begin{equation}
{\mathbb S}_i=-\frac{1}{k}\widehat{D}_i{\mathbb S},
\end{equation}
\begin{equation}
{\mathbb S}_{ij}=\frac{1}{k^2}\widehat{D}_i\widehat{D}_j{\mathbb S}
+\frac1n\gamma_{ij}{\mathbb S},
\end{equation}
where $\widehat{D}_i$ denotes the covariant derivative on the unit sphere 
and $k$ is defined in Eq.~\eqref{def-kk}.
Since the initial metric is the Brill-Lindquist one,
the values of $H_2, H_L, H_T, h_1$ are same as the ones that we 
have derived in \cite{YSS05}:
\begin{equation}
H_2=2H_L=\chi(r)\equiv\frac{1/(n-1)R^{n-1}}{1+1/4R^{n-1}}
\left(\frac{z_0}{R}\right)^2\left(K_2^{[n]}\right)^{-1},
\label{chi}
\end{equation}
\begin{equation}
h_1=H_T=0. 
\end{equation}
Here our purpose is to compute the time derivative of these variables.

The nonzero components of $\widehat{K}_{\mu\nu}$ shown in 
Eqs.~\eqref{Krr-close}--\eqref{Kpp-close} are rewritten as
\begin{equation}
\widehat{K}_{RR}=-(z_0P/M_0)
\left[(n+2)\frac{S_2^{[n]}}{K_2^{[n]}}+n\frac{S_0^{[n]}}{K_0^{[n]}}\right]R^{-(n+1)},
\end{equation}
\begin{equation}
\widehat{K}_{ij}
=(z_0P/M_0)
\left[
\frac{n+2}{n}\frac{S_2^{[n]}}{K_2^{[n]}}\gamma_{ij}
+\frac{2(n+1)}{n-1}\frac{S_{2~ij}^{[n]}}{K_2^{[n]}}
+\frac{S_0^{[n]}}{K_0^{[n]}}\gamma_{ij}
\right]R^{-(n-1)}.
\end{equation}
Thus $K_{\mu\nu}$ is composed of $\ell=0$ and $2$ modes. Since 
$\ell=0$ mode is absorbed in a coordinate transformation,
we omit it hereafter. The extrinsic curvature 
$K_{\mu\nu}=\varPsi^{-2}\widehat{K}_{\mu\nu}\simeq\varPsi_0^{-2}
\widehat{K}_{\mu\nu}$ [where $\varPsi_0$ is defined in Eq.~\eqref{hat-Psi-0}] 
in the $(r,\theta,\phi_*)$ coordinate is
\begin{equation}
K_{rr}=-(z_0P/M_0)\left[(n+2)\frac{S_2^{[n]}}{K_2^{[n]}}\right]
f^{-1}\varPsi_0^{-2(n+1)/(n-1)}R^{-(n+1)}.
\label{Krr}
\end{equation}
\begin{equation}
K_{ij}=(z_0P/M_0)\left[
\frac{n+2}{n}\frac{S_2^{[n]}}{K_2^{[n]}}\gamma_{ij}
+\frac{2(n+1)}{n-1}\frac{S_{2~ij}^{[n]}}{K_2^{[n]}}
\right]\varPsi_0^{-2}R^{-(n-1)}.
\label{Kij}
\end{equation}

$K_{\mu\nu}$ is written as
\begin{equation}
K_{\mu\nu}=-\frac{1}{2\alpha}
\left(\dot{h}_{\mu\nu}-\nabla_\mu \beta_\nu-\nabla_\nu \beta_\mu\right)
\end{equation}
in terms of the lapse function $\alpha=\sqrt{f}$
and the shift vector $\beta_\mu$.
Although $\beta_\mu$ can be freely chosen provided that $O(\beta_\mu)=O(z_0^2)$,
we impose $\beta_\mu=0$ for simplicity. This brings us $H_1=h_0=0$
and $\dot{h}_{\mu\nu}=-2\sqrt{f}K_{\mu\nu}$. Using Eqs.~\eqref{Krr} and \eqref{Kij},
we have the explicit formulas of $\dot{h}_{rr}$
 and $\dot{h}_{ij}$ and they are compared with Eq.~\eqref{eq:scalar-perturbation}:
\begin{equation}
\dot{h}_1=0,
\end{equation}  
\begin{equation}
\dot{H}_2=2(n+2)\eta(r),
\end{equation}
\begin{equation}
\dot{H}_L=-\frac{n+2}{n}\eta(r),
\end{equation}
\begin{equation}
\dot{H}_T=-\frac{2(n+1)}{n-1}\eta(r),
\label{HT}
\end{equation}
where
\begin{equation}
\eta(r)=\frac{(z_0P/M_0)}{K_2^{[n]}}
\frac{\sqrt{f}}{r^{n+1}}.
\label{eq:eta}
\end{equation}

Now we are in a position to derive the equation for $\Phi$ and $\dot{\Phi}$.
Setting
\begin{equation}
f_{ab}=\left(
\begin{array}{cc}
fH_0 & H_1\\
H_1 & H_2/f
\end{array}
\right),~~
rf_a=\left(h_0, h_1\right),
\end{equation}
the gauge invariant quantities derived in \cite{KI03} are
\begin{equation}
F= H_L+(1/n)H_T+(1/r)D^arX_a,
\end{equation}
\begin{equation}
F_{ab}=f_{ab}+D_aX_b+D_bX_a,
\end{equation}
where
\begin{equation}
X_a=\frac{r}{k}\left(f_a+\frac{r}{k}D_aH_T\right),
\end{equation}
and $D_a$ denotes the covariant derivative with respect to the
metric $g_{ab}dy^ady^b=-fdt^2+f^{-1}dr^2$.
In our case, 
\begin{equation}
X_t=\frac{r^2}{k^2}\dot{H}_T,~~X_r=0,
\end{equation}
and
\begin{equation}
F=\chi(r)/2.
\end{equation}
$F$ is related to $\Phi$ and $\Phi_{,r}$ as shown in \cite{KI03}
and we find the same equation for $\Phi$ as that 
in our previous paper \cite{YSS05}. Hence, the initial value of $\Phi$ is unchanged.
Calculating $F_{rt}$, we obtain 
\begin{equation}
F_{rt}=-\frac{2(n+1)}{n-1}
\frac{r^2}{k^2}
\left[
\left(\frac{2}{r}-\frac{f_{,r}}{f}\right)\eta(r)+2\eta_{,r}(r)
\right],
\end{equation}
where we have substituted Eq.~\eqref{HT}.
Since $F_{rt}$ is related to $\dot{\Phi}$ and $\dot{\Phi}_{,r}$
as
\begin{equation}
F_{rt}=r^{1-n/2}\left(-\frac{P_Z}{4Hf}\dot{\Phi}+r\dot{\Phi}_{,r}\right),
\end{equation}
we find the equation
\begin{equation}
\dot{\Phi}_{,r}
=\frac{P_Z}{4Hrf}\dot{\Phi}
+\frac{2n}{n-1}r^{n/2-1}\eta(r),
\label{eq:initial-master-dot-equation}
\end{equation}
where $P_Z$ is a function of $r$ given in \cite{KI03}
and we used Eq.~\eqref{eq:eta}.
The solution is given by Eq.~\eqref{initial-dot-Phi}.

\end{document}